\documentclass[%
 reprint,
superscriptaddress,
showpacs,preprintnumbers,
amsmath,amssymb,
aps,
prl,
]{revtex4-1}

\usepackage{graphicx}
\usepackage{dcolumn}
\usepackage{bm}
\usepackage{multirow}

\usepackage{amssymb}
\usepackage{amsmath}
\usepackage{colordvi}
\usepackage[colorlinks]{hyperref}
\usepackage{amsthm}
\usepackage{subeqnarray}
\usepackage{cases}
\usepackage{enumerate}
\usepackage{graphicx}
\usepackage{epsfig}
\usepackage{epstopdf}
\usepackage{subfigure}
\usepackage{color}
\usepackage{bm}

\def\avg(#1){\langle#1\rangle}

\def\be{\begin{equation}}
\def\ee{\end{equation}}
\def\bea{\begin{eqnarray}}
\def\eea{\end{eqnarray}}

\usepackage{ulem}
\newcommand{\red}[1]{{\color{red}#1}}

\begin{document}


\title{Super-Heisenberg scaling in a triple point criticality}

\author{Jia-Ming Cheng}
\affiliation{Xi'an Microelectronics Technology Institute, Xi'an, 710065, P.R. China}

\author{Yong-Chang Zhang}
\email{zhangyc@xjtu.edu.cn}
\affiliation{MOE Key Laboratory for Nonequilibrium Synthesis and Modulation of Condensed Matter,
Shaanxi Key Laboratory of Quantum Information and Quantum Optoelectronic Devices,
School of Physics, Xi'an Jiaotong University, Xi'an 710049, P.R. China}

\author{Xiang-Fa Zhou}
\email{xfzhou@ustc.edu.cn}
\affiliation{CAS Key Lab of Quantum Information, University of Science and Technology of China, Hefei, 230026, P.R. China}
\affiliation{Synergetic Innovation Center of Quantum Information and Quantum Physics, University of Science and Technology of China, Hefei, 230026, P.R. China}

\author{Zheng-Wei Zhou}
\email{zwzhou@ustc.edu.cn}
\affiliation{CAS Key Lab of Quantum Information, University of Science and Technology of China, Hefei, 230026, P.R. China}
\affiliation{Synergetic Innovation Center of Quantum Information and Quantum Physics, University of Science and Technology of China, Hefei, 230026, P.R. China}

\date{\today}

\begin{abstract}
	We investigate quantum-enhanced metrology in a triple point criticality and discover that quantum criticality does not always enhance measurement precision.
	We have developed suitable adiabatic evolution protocols to effectively restrain excitations, which could accelerate the adiabatic evolutions and lead to an exponential super-Heisenberg scaling.
	This scaling behavior is quite valuable in practical parameter estimating experiments with limited coherence time.
	Dissipation can strengthen the super-Heisenberg scaling until decoherence increases to dominate in the dissipative dynamics.
	Additionally, measurement precisions beyond Heisenberg scaling can be experimentally achieved in the trapped ion system.
	Our findings strongly indicate that criticality-enhanced metrology can indeed significantly enhance measurement precisions to a super-Heisenberg scaling when combining a triple point and beneficial parameter modulations, which will be conducive to the exploration of other super-Heisenberg scaling and their applications.
\end{abstract}

\maketitle


\textit{Introduction.}---Quantum metrology focuses on improving measurement precision by exploiting quantum resources, such as entanglement and squeezing \cite{pezze2018quantum}, and has significant impacts on developments of fundamental sciences \cite{braun2018quantum}, which has undoubtedly been attracting widespread attentions \cite{wang2019heisenberg,qin2023unconditional,lu2024critical,giovannetti2004quantum,yang2022variational,len2022quantum,chu2023strong,salvia2023critical}.
Currently, primary aims of quantum metrology are beating the quantum standard limit (QSL) and saturating the Heisenberg limit (HL) \cite{ilias2024criticality,hou2021super,cimini2023experimental}.
The achievable precision limit is tightly depended on quantum resources, such as the number of independent probes $\mathcal{N}$ and total duration $\mathcal{T}$.
For classical measurements, the optimal estimating error should be the QSL: $\sim1/\sqrt{\mathcal{NT}}$ \cite{escher2011general,brask2015improved,mason2019continuous}.
When estimating involves quantum correlations, its standard error should satisfy the HL: $\sim1/(\mathcal{NT})$ \cite{giovannetti2006quantum,daryanoosh2018experimental,zhou2018achieving}.

Such quantum enhancement beyond the QSL in estimating accuracy can be witnessed by harnessing entangled states, e.g., Greenberger-Horne-Zeilinger states \cite{bollinger1996optimal,zhao2021creation,frantzeskakis2023extracting}, NOON states \cite{dowling2008quantum,slussarenko2017unconditional,afek2010high}, and squeezed states \cite{kitagawa1993squeezed,davis2016approaching,sinatra2022spin}.
Preparations of these states with a large number of constituents are technically challenging. They are particularly susceptible to external noises \cite{huelga1997improvement}, which limits their applications in real quantum-enhanced metrology \cite{marciniak2022optimal,song2019generation,omran2019generation}.
Besides entanglement, quantum criticality has been confirmed as an alternative resource \cite{zanardi2008quantum,di2023critical,garbe2022critical,gietka2021adiabatic,gietka2022understanding,gietka2022squeezing,gietka2023squeezing}. The crucial ingredient of criticality-enhanced metrology lies in that the susceptibility of ground states at a critical point is divergent \cite{gu2009scaling,cheng2017scaling}, and it becomes extremely sensitive to even tiny variations in the underlying Hamiltonian \cite{sachdev1999quantum}.
Adiabatically driving a physical system to the vicinity of a critical point, measurement precision could be raised to the HL \cite{garbe2020critical,invernizzi2008optimal,rams2018limits,ilias2022criticality}.
Such a requirement of adiabatic evolutions in preparing a ground state makes it difficult to fulfill the criticality-enhanced metrology.

Beyond HL, a super-Heisenberg scaling (HS) is regarded to emerge in a parameterized Hamiltonian including $k$-body nonlinear interaction, which scales as $1/\mathcal{N}^k$ with $k>1$ \cite{boixo2007generalized,napolitano2011interaction,hall2012does}.
However, this many-body interaction is not easy to create.
As the quantum Fisher information (QFI) is usually divergent at a critical point, parameter estimations could be arbitrarily accurate resulting in an apparent super-HS.
It has been shown that this super-HS would relax to the HS when taking into account the time needed to accomplish adiabatic evolutions \cite{rams2018limits}.
To adiabatically follow the instantaneous ground states, the slow ramp rate $v$ is often set to be much smaller than energy gap $\Delta$, for example, $v\approx\delta\Delta^3/\omega_0^2$ with $\delta\ll1$ and characteristic frequency $\omega_0$ ($\hbar=1$) \cite{garbe2020critical}.
It nearly becomes a scientific consensus that HL is the ultimate precision for a criticality-enhanced metrology \cite{rams2018limits,ilias2022criticality}.
Two questions naturally arise: Can quantum criticality always improve measurement precision? 
Whether a super-HS can appear if the slow ramp rates can be set faster but still much smaller than the energy gap, $v\approx\delta\Delta$, for instance?

In this work, we solve both queries convincingly by taking the anisotropic quantum Rabi model (aQRM) as an example. 
We devise adiabatic modulating protocols that can restrain excitations and speed the adiabatic process, find out analytical forms of the QFI and figure out an exponential super-HS.
Dissipation of spin decay and photon loss can enhance the super-HS before decoherence in the dissipative dynamics starts to destroy it.
Divergence of the QFI will lose in some other parameter modulating schemes, demonstrating that quantum criticality can not always enhance measurement precision.
We also suggest a feasible experimental scheme to perform precise estimations beyond the HS.

\textit{QFI in quantum  criticality.}---As QFI plays a vital role in parameter estimations, we initially study characteristics of QFI in quantum criticality.
The general Hamiltonian of a quantum system experiencing quantum phase transitions can be expressed as $H(\lambda)=H_0+\lambda H_1$. $H_1$ is supposed to be the driving term with controlling parameter $\lambda$.
It is assumed that this Hamiltonian has eigenvalues $E_n(\lambda)$ and eigenstates $\vert\psi_n(\lambda)\rangle$ with $n=0,1,2,\cdots$. $E_0(\lambda)$ is its ground energy.
$\lambda=\lambda_c$ represents the critical point. We further assume that ground state has no degeneracy. 
The information about $\lambda$ that can be extracted from this quantum criticality are restricted by QFI relative to $\lambda$. By non-degenerate perturbed theory \cite{sakurai2020modern}, its form should read as [see Supplemental Material (SM) \red{I}]
\begin{align}
	F_\lambda=4\sum_{n\ne0}\frac{\vert\langle\psi_n(\lambda)\vert H_1\vert\psi_0(\lambda)\rangle\vert^2}{[E_0(\lambda)-E_n(\lambda)]^2},
\end{align}
from which we can learn that:
\begin{enumerate}
	\item As it is well known that around a normal critical-point of first order or second order, QFI will be divergent because the energy gap closes: $E_0(\lambda)-E_n(\lambda)=0$, and meantime it is usually true that $\langle\psi_n(\lambda)\vert H_1\vert\psi_0(\lambda)\rangle\ne0$, which construct the basis of quantum criticality-enhanced metrology \cite{yang2007ground,gu2008fidelity,gu2010fidelity}. 
		In this criticality, excitations in an adiabatic evolution can not be restrained [see SM \red{I}], so the slow ramp rates must be much smaller than the corresponding energy gaps, leading to the renowned HS \cite{garbe2020critical,rams2018limits} [see SM \red{V}];
	\item However, around a critical point, such as a triple point, when both terms $E_0(\lambda)-E_n(\lambda)$ and $\langle\psi_n(\lambda)\vert H_1\vert\psi_0(\lambda)\rangle$ approach zero simultaneously, then QFI will be either finite or divergent, depending on their relative speed of approaching zero. Such a mechanism provides us a way to modulate the QFI.
		This criticality may allow us to suppress excitations in the adiabatic evolutions and set larger slow ramp rates [see SM \red{I}], which can ultimately reduce the evolution times and achieve a super-HS.
		If the QFI is finite, it would not be useful for quantum metrology.
\end{enumerate}

\begin{figure}[htpb]
	\centering
	\includegraphics[width=0.43\textwidth]{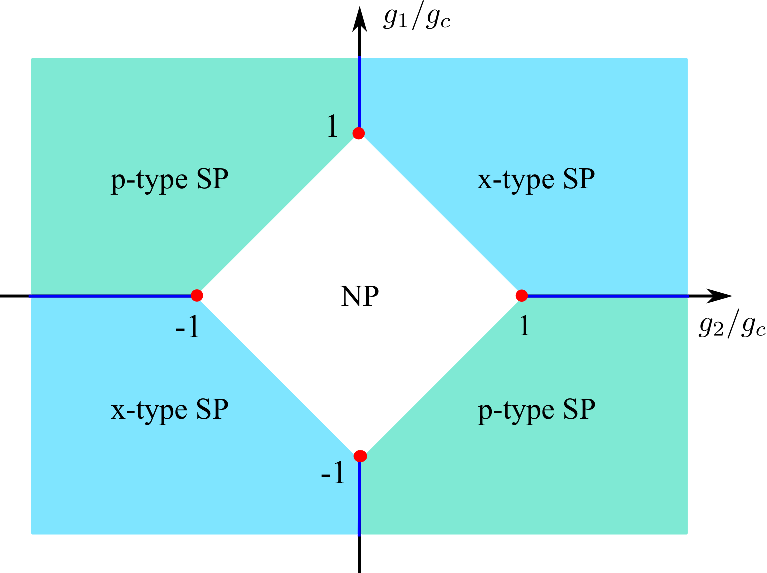}
	\caption{\label{fig:f0}
	Phase diagram of the aQRM. ``NP" and ``SP" are short for normal phase and superradiant phase respectively. Phase transitions from NP to SP are continuous spontaneously breaking the parity symmetry, while those between SPs turn out to be discontinuous.
	There exist four triple points marked by red dots: $(g_1,g_2)/g_c=(\pm 1,0)$, $(0,\pm 1)$ with $g_c=2\sqrt{\omega\Omega}$.
	}
\end{figure}

\textit{Critical metrology based on aQRM.}---We consider criticality-enhanced quantum metrology around a triple point in the aQRM, whose Hamiltonian is ($\hbar=1$)
\begin{align}
	\label{eq:ham}
	H=\frac{\Omega}{2}\sigma_z+\omega a^\dagger a+(\frac{g_1}{2}a^\dagger\sigma_-+\frac{g_2}{2}a^\dagger\sigma_++\text{h.c.}),
\end{align}
where $\sigma_{z,\pm}$ are Pauli operators of the two-level system with transition frequency $\Omega$, ground state $\vert\downarrow\rangle$ and excited state $\vert\uparrow\rangle$, $a^\dagger$ ($a$) is creation (annihilation) operator of the light field with frequency $\omega$, $g_1$ and $g_2$ are their rotating-wave and counter-rotating-wave coupling strengths.
This Hamiltonian possesses parity symmetry with symmetric operator $\mathcal{P}=\exp[i\pi(a^\dagger a+\sigma_z/2)]$. 
In the infinite frequency ratio limit $\Omega/\omega\to+\infty$, this system would undergo quantum phase transitions \cite{liu2017universal}.
Triple points locate at the cross points of its phase boundaries [see FIG. \ref{fig:f0} and SM \red{II}].
As transition frequency of the two-level system is dominated, 
it tends to stay at its ground state $\vert\downarrow\rangle$.
We are interested in low-energy physics of the light-field in weak interactions. Applying a Schrieffer-Wolff transformation $U_n=\exp[\frac{g_1a\sigma_+}{2(\Omega-\omega)}+\frac{g_2a^\dagger\sigma_+}{2(\Omega+\omega)}-\text{h.c.}]$ to Eq. (\ref{eq:ham}) and then projecting to the subspace of $\vert\downarrow\rangle$, the effective Hamiltonian is proved to be
\begin{align}
	H_{np}=\omega[ (1-\frac{g_1^2+g_2^2}{g^2_c})a^\dagger a-\frac{g_1g_2}{g^2_c}(a^{\dagger 2}+a^2) ]
\end{align}
to second order of interactions $g_1$ and $g_2$, where we have ignored the constant terms.
Hamiltonian $H_{np}$ can be diagonalized by a squeezing transformation $\Gamma(\gamma)=\exp[\frac{\gamma}{2}(a^2-a^{\dagger 2})]$ with $\gamma=\frac{1}{4}\ln\frac{g^2_c-(g_1+g_2)^2}{g^2_c-(g_1-g_2)^2}$, resulting in the energy gap $\Delta=\omega[1-(\frac{g_1-g_2}{g_c})^2]^{1/2}[1-(\frac{g_1+g_2}{g_c})^2]^{1/2}$ and ground state $\vert\psi_{np}\rangle=\Gamma(\gamma)\vert0\rangle$ of NP.

\begin{figure}[htpb]
	\centering
	\includegraphics[width=0.43\textwidth]{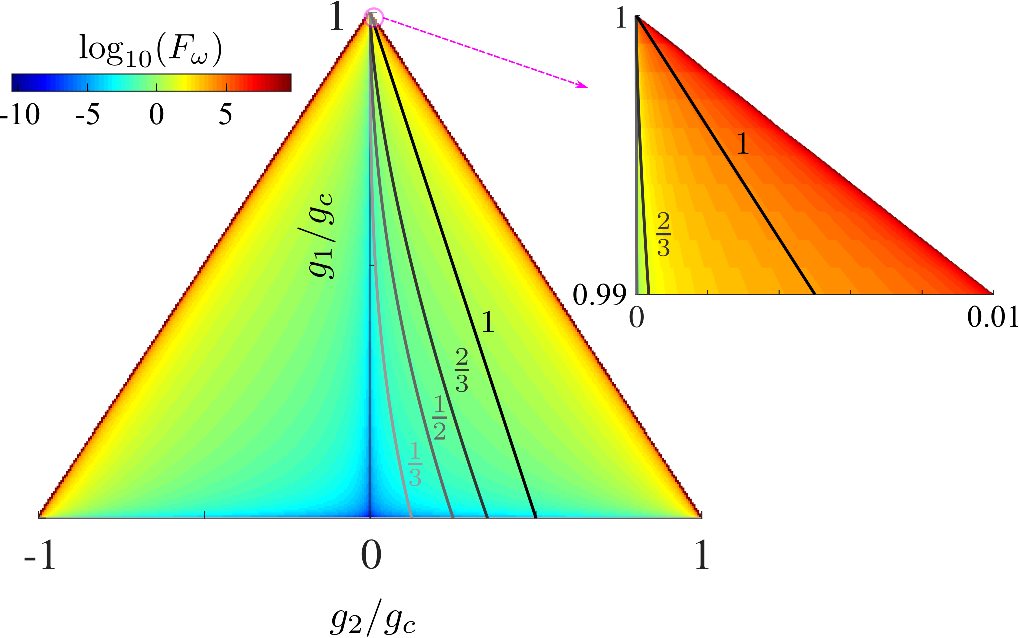}
	\caption{\label{fig:f1}
	QFI $F_\omega$ in the NP and adiabatic evolution paths $\frac{g_1}{g_c}=1-2(\frac{g_2}{g_c})^\beta$ to final points near the triple point $(g_1,g_2)=(g_c,0)$ with $g_c=500$, $\Omega/\omega=10^6$, $\beta=\frac{1}{3}$, $\frac{1}{2}$, $\frac{2}{3}$ and $1$.
	}
\end{figure}

In order to accomplish criticality-enhanced parameter estimations, we first prepare the aQRM in its ground states far from the phase boundaries, then gradually bring the system in close proximity to a critical point and finally measure relevant observables.
In particular, we show an instance of measuring the light-field frequency $\omega$.
Its estimating precision is bounded by the quantum Cram\'er-Rao bound (QCRB): $\Delta\omega\ge1/\sqrt{\nu F_\omega}$ \cite{escher2011quantum}, where $\nu$ is the number of independent measurements and $F_\omega$ is the QFI relative to parameter $\omega$.
Since the final point of adiabatic evolutions is near a phase boundary and its corresponding ground state is in the NP, the QFI can be computed exactly as $F_\omega=4[ \langle\partial_\omega\psi_{np}\vert\partial_\omega\psi_{np}\rangle-\vert\langle\partial_\omega\psi_{np}\vert\psi_{np}\rangle\vert^2 ]=2(\partial_\omega\gamma)^2$, whose concrete form becomes
\begin{align}
	\label{eq:fomega}
	F_\omega=\frac{1}{8\omega^2}[\frac{1}{1-(\frac{g_1+g_2}{g_c})^2}-\frac{1}{1-(\frac{g_1-g_2}{g_c})^2}]^2.
\end{align}
It is evident that at a phase boundary $\vert g_1+g_2\vert=g_c$ or $\vert g_1-g_2\vert=g_c$, the QFI definitely diverges [see FIG. \ref{fig:f1}]. 
Nevertheless, the adiabatic evolution time $T$ will also lengthen to infinity accompanying with closed energy gaps, which is known as critical slowing down.
However, near a triple point,  whether the QFI is divergent depends on the specific relations between $g_1$ and $g_2$.
For example, around the triple point $(g_1,g_2)=(g_c,0)$, 
we assume $1-\frac{g_1}{g_c}=k(\frac{g_2}{g_c})^\beta (k>1, 0<\beta\le1)$,
then characteristics of the QFI $F_\omega$ can be found in TABLE \ref{tab:t0}.
Here appropriate power exponent $\beta$ and coefficient $k$ should be selected to ensure that the evolution path is located in the NP and far away from the phase boundaries.
It clearly manifests that divergent behaviors of the QFI can be modulated by changing $g_1$ and $g_2$ [see FIG. \ref{fig:f1}], which also indicates that the excitations in an adiabatic process may be partially restrained, and 
so the slow ramp rates could be faster but still much smaller than the energy gap,
overcoming the critical slowing down effects and achieving a super-HS.

\begin{table}[b]
\caption{\label{tab:t0}%
Varying of the QFI $F_{\omega}$ along with power exponent $\beta$ in case of $1-\frac{g_1}{g_c}=k(\frac{g_2}{g_c})^\beta (k>1, 0<\beta\le1)$.
}
\begin{ruledtabular}
\begin{tabular}{lcr}
\textrm{$\beta$}&
\textrm{$F_\omega$}&
\textrm{finite/divergent}\\
\colrule
	$0<\beta<\frac{1}{2}$ & $\frac{1}{8}\omega^{-2}k^{-4}(\frac{g_2}{g_c})^{2(1-2\beta)}$ & $F_\omega\to0$ \\
	$\beta=\frac{1}{2}$ &  $\frac{1}{8}\omega^{-2}k^{-4}$ & finite \\
	$\frac{1}{2}<\beta<1$ &  $\frac{1}{8}\omega^{-2}k^{-4}(\frac{g_2}{g_c})^{2(1-2\beta)}$ & divergent \\
	$\beta=1$ & $\frac{1}{8}\omega^{-2}(k^2-1)^{-2}(\frac{g_2}{g_c})^{-2}$ & divergent \\
\end{tabular}
\end{ruledtabular}
\end{table}

\begin{figure}[htpb]
	\centering
	\includegraphics[width=0.47\textwidth]{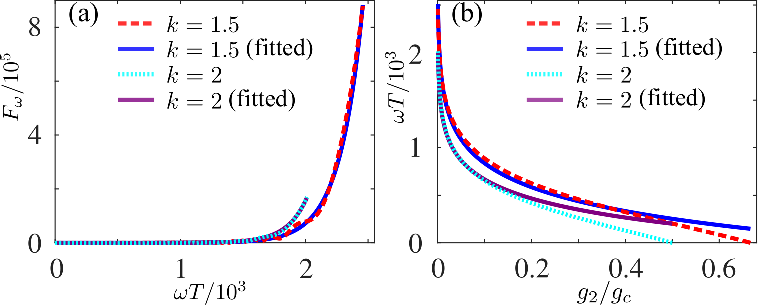}
	\caption{\label{fig:f2}
	QFI (a) and adiabatic evolution time (b) of 
	two evolution paths 
	to final points near triple point $(g_1,g_2)=(g_c,0)$ with $\Omega/\omega=10^6$, $g_c=500$, $\delta=10^{-3}$.
	In both figures, the red-dashed line and aqua-dashed line stand for results of real-time adiabatic evolutions with rates $k=1.5,2$, the blue-solid line and purple-solid line are their corresponding fitted results.
	The fitted function of QFI is $F_\omega=ae^{bT}$ with $a=1.3\times10^{-2}$, $b=5.5\times 10^{-3}$ for $k=1.5$ and $a=5.5\times10^{-3}$, $b=1.6\times 10^{-3}$ for $k=2$.
	The fitted function of time is $T=-a\ln\frac{g_2}{g_c}$ with $a=1455.2$ if $k=1.5$ and $a=1159.2$ if $k=2$.
	We vary $g_2$ form $g_c/k$ to the final value $0.001g_c$. 
	}
\end{figure}

\textit{Super-Heisenberg scaling.}---Using a general adaptive manner, we adapt adiabatically interactions $g_1$ and $g_2$ along the straight line: $g_1(t)+kg_2(t)=g_c$ with $g_2(t)/g_c=1/k-\int^t_0 v(t^\prime)dt^\prime$ [see SM \red{IV.A}].
\textcolor{red}{
	We start at the initial point $(g^{(i)}_1,g^{(i)}_2)=(0,g_c/k)$, whose corresponding ground state is a vacuum of the light field.
}
When approaching the triple point $(g_1,g_2)=(g_c,0)$, the QFI is approximated as $F_\omega\approx(\frac{g_2}{g_c})^{-2}/[8\omega^2(k^2-1)^2]$, whose divergent behaviors are partially countervailed by a large rate $k$, apparently [see FIG. \ref{fig:f2}(a)].
Thus a small rate $k$ is preferred to reach high measurement precisions.
Based on adiabatic evolution theory [see SM \red{III}], the probability of excitations is directly proportional to $\partial_{g_2}\gamma$, which can be written as
\begin{align}
	\label{eq:partgamma}
	\frac{\partial\gamma}{\partial g_2}=\frac{(g_1-g_2)(\frac{\partial g_1}{\partial g_2}-1)}{2[g^2_c-(g_1-g_2)^2]}-\frac{(g_1+g_2)(\frac{\partial g_1}{\partial g_2}+1)}{2[g^2_c-(g_1+g_2)^2]},
\end{align}
and is divergent around a phase boundary, such as, $g_1+g_2=g_c$. So the adaptive slow ramp rate $v(g_2)$ must be set small enough to guarantee an adiabatic process. 
Whereas close to the triple point $(g_1,g_2)=(g_c,0)$, the situations are quite different, similar to that of the QFI, its divergency also relies on the relationships between $g_1$ and $g_2$,
for example, if we have $1-\frac{g_1}{g_c}=k(\frac{g_2}{g_c})^{1/2}$, then $\partial_{g_2}\gamma\approx-\frac{1}{4kg_c}(\frac{g_2}{g_c})^{-1/2}$ is divergent, which means that the excitations can not be restrained;
while in our adiabatic modulations, $\partial_{g_2}\gamma=\frac{1}{g_c}[2-(k-1)\frac{g_2}{g_c}]^{-1}[2-(k+1)\frac{g_2}{g_c}]^{-1}\approx\frac{1}{4g_c}$ is finite,
which reveals that our modulating scheme can suppress excitations and therefore accelerate the ground state evolution. 

To assess the performances of this adiabatic protocol, it is necessary to consider the evolution time $T$ and average photon number $N$ that are used in the critical sensing.
At the final points, average photon number $N=\langle\psi_{np}\vert n\vert\psi_{np}\rangle\approx\frac{k}{2\sqrt{k^2-1}}-\frac{1}{2}$, here $n=a^\dagger a$.
Making use of time-dependent perturbation theory, the optimal slow ramp rate could be determined as $v(g_2)=\frac{2\delta}{k}\Delta(g_2)\approx\frac{4\delta\omega}{k}(k^2-1)^{1/2}\frac{g_2}{g_c}$ with $\delta\ll1$.
Hence, the evolution time is given by 
$T=\int^{g_2}_{g_c/k}\frac{1}{-g_cv(g)}dg\approx -\frac{1}{4\delta\omega}(1-\frac{1}{k^2})^{-1/2}\ln\frac{g_2}{g_c}$
and a big rate $k$ is required to shorten it [see FIG. \ref{fig:f2}(b)]. 
As a result, we obtain a super-HS with respect to time $T$
\begin{align}
	\label{eq:fot}
	F_\omega\approx&\frac{1}{8\omega^2(k^2-1)^2}e^{8\delta\omega\sqrt{1-\frac{1}{k^2}}T},
\end{align}
from which it can be known that the sensing protocol suggested around a triple point can greatly surpass the HS with respect to time $T$.
This exponential scaling can make it possible to overcome the dilemma of finite coherence time in actual critical metrology.
 An appropriate rate $k$ should be selected to pursue a large QFI $F_\omega$ within a relatively short time $T$.
For more complex modulating paths similar super-HS will appear if they approach the triple point in the same manner as that of the straight line [see SM \red{IV.B}].
Similarly, the exponential super-HS can also appear in the quantum Jaynes-Cummings model (JCM) with a squeezing bosonic mode [see SM \red{VIII}]. 
If an adiabatic modulation approaching the triple point fails to restrain excitations, i.e., $\partial_{g_2}\gamma$ is not finite but divergent, we would arrive at a sub-HS [see SM \red{IV.C}].
The super-HS is enabled by combining a triple point criticality and suitable parameter modulations that can weaken the critical slowing down.

\begin{figure}[htpb]
	\centering
	\includegraphics[width=0.47\textwidth]{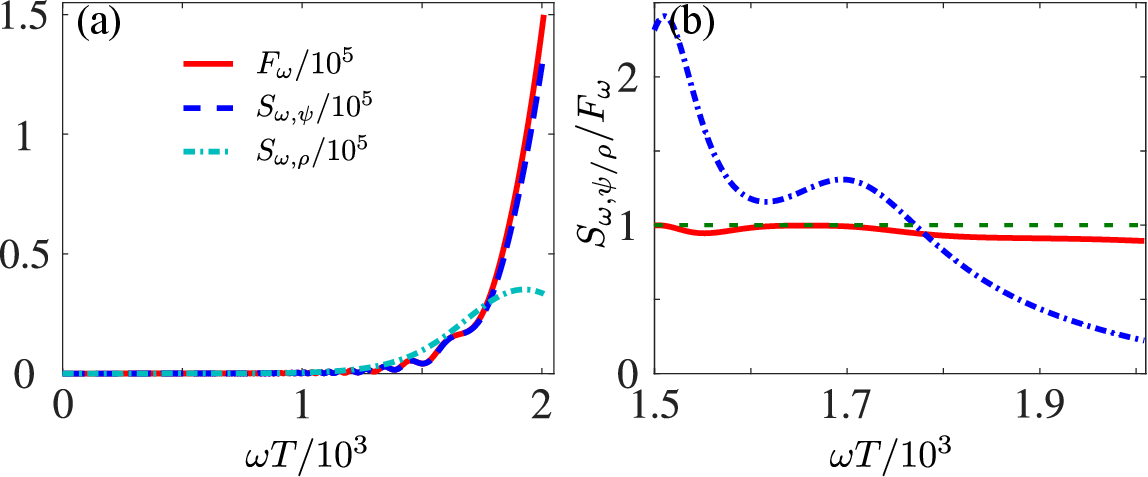}
	\caption{\label{fig:f3}
	(a) QFI $F_\omega$ or SNR $S_{\omega,\psi/\rho}$ v.s. adiabatic evolution time $T$, and (b) ratios $S_{\omega,\psi}/F_\omega$ (red line) and $S_{\omega,\rho}/F_\omega$ (blue line) in the adiabatic evolution protocol to a final point near triple point $(g_1,g_2)=(g_c,0)$ with parameters $\Omega/\omega=10^6$, $g_c=500$, $\delta=10^{-3}$ and $k=2$.
	We vary $g_2$ from $g_c/k$ to the final value $0.001g_c$. 
	When including dissipation, we set $\mathcal{K}_p=0.01\omega$ and $\mathcal{K}_a=0.01\Omega$.
	}
\end{figure}

To saturate the QCRB, we choose to measure photon number $\langle n\rangle$ in a final state. The measurement precision of light-field frequency is determined by $\Delta\omega=S_{\omega,\psi}^{-1/2}$ with singal-to-noise ratio (SNR) $S_{\omega,\psi}=(\partial_\omega\langle n\rangle)^2/(\langle n^2\rangle-\langle n\rangle^2)$ [see SM \red{VI}].
In the NP, we can verify that $S_{\omega,\psi}=2(\partial_\omega\gamma)^2=F_\omega$. In FIG. \ref{fig:f3}, it is shown that $S_{\omega,\psi}\approx F_{\omega}$ during adiabatic evolutions around the triple point.

\textit{Effects of dissipation.}---To analyze the effects of dissipation in open aQRM due to photon loss and spin decay, we consider dissipative dynamics described by a Lindblad master equation
\begin{align}
	\frac{\partial}{\partial t}\rho(t)=-i[H,\rho(t)]+\mathcal{K}_{p}\mathcal{L}[a]\rho(t)+\mathcal{K}_{a}\mathcal{L}[\sigma_-]\rho(t)
\end{align}
with dissipative rates $\mathcal{K}_p\ll\omega$, $\mathcal{K}_a\ll\Omega$ and damping superoperator $\mathcal{L}[O]\rho=O\rho O^\dagger-\frac{1}{2}\{O^\dagger O,\rho\}$.
After inclusion of dissipation, decoherence will accumulate in the previous adiabatic evolutions and the final state would become incoherent and mixed over a long time evolution, which is illustrated in FIG. \ref{fig:f3}(a) that SNR $S_{\omega,\rho}$ first rises then descends.
It is surprising that this SNR $S_{\omega,\rho}$ could exceed the QFI $F_\omega$ in a period of time, which reveals that dissipation can raise the measurement precisions.
This enhancement may be attributed to the non-phase-covariant noises, which can enhance quantum metrology due to non-commutation between coherent and dissipative dynamics \cite{peng2024enhanced}.

\begin{figure}[htpb]
	\centering
	\includegraphics[width=0.48\textwidth]{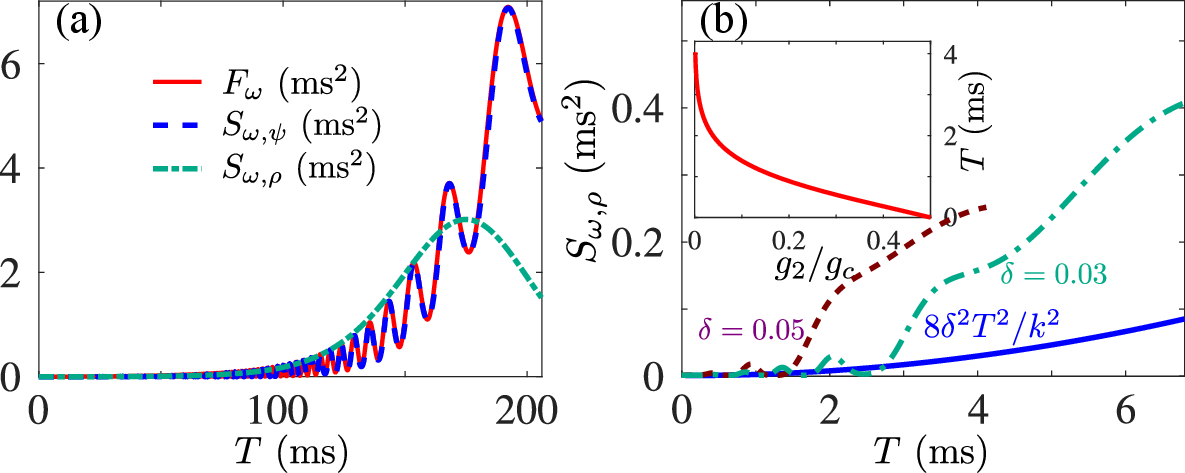}
	\caption{\label{fig:f4}
	QFI $F_\omega$ or SNR $S_{\omega,\psi/\rho}$ v.s. time $T$ in the evolution protocol to a final point near triple point $(g_1,g_2)=(g_c,0)$ with $k=2$, $\omega=2\pi\times1$ KHz, $\Omega=2\pi\times250$ KHz.
	In (a), $\delta=0.001$ and in the subfigure of (b), $\delta=0.05$.
	We vary $g_2$ form $g_c/k$ to the final value $0.001g_c$. 
	When including dissipation, we set $\mathcal{K}_p=0.01\omega$ and $\mathcal{K}_a=0.01\Omega$.
	}
\end{figure}

\textit{Experimental feasibility.}---Up to now, experimental implementations of criticality-enhanced quantum metrology have been believed to be almost unfeasible due to the intrinsic critical slowing down \cite{liu2021experimental}.
However, our adiabatically modulating protocols designed around a triple point resulting in the exponential scaling will expend much less time and thus could potentially surmount the impediment of limited coherence time in an actual system.
With a view to experimental realizations of simulating the QRM and observing its phase transitions with a single trapped $^{171}\text{Yb}^+$ ion \cite{lv2018quantum,cai2021observation},
we advise carrying out this experiment using its two hyperfine states from the ground-state manifold $^{2}$S$_{1/2}$, i.e., $\vert\uparrow\rangle=\vert F=1,m_F=0\rangle$ and $\vert\downarrow\rangle=\vert F=0,m_F=0\rangle$ with transition frequency $\omega_q\approx 2\pi\times 12.6$ GHz.
The spatial motion along its principal axes $x$ is cooled close to the ground state with frequency $\omega_x=2\pi\times5 $ MHz \cite{leibfried2003quantum},
which can be well described as a quantum harmonic oscillator and serves as the bosonic mode (light-field).  
Employing the experimental scheme in \cite{cai2021observation}, but with different Rabi frequencies $\Omega_{b,r}$, we arrive at an effective interaction picture Hamiltonian of the aQRM as presented in Eq. (\ref{eq:ham}) with $\Omega=(\delta_b+\delta_r)/2$, $\omega=(\delta_b-\delta_r)/2$ and $g_{1,2}=\eta_{r,b}\Omega_{r,b}$.
$\delta_{b,r}\ll\omega_x$ are detuning of the blue- or red-sidebands \cite{lv2018quantum}. 
We fix the detuning as $\delta_b=2\pi\times251$ KHz, $\delta_r=2\pi\times249$ KHz. 
As shown in FIG. \ref{fig:f4}(a), we acquire a largest SNR $S_{\omega,\rho}$ at about $175$ ms, which is much longer than the motional coherence time of about $5$ ms \cite{cai2021observation}.
The highest precision is $\Delta\omega\approx0.092\omega$. Because of the small available frequency-ratio $\Omega/\omega$, we can not reach the super-HS scaling in Eq. (\ref{eq:fot}).
However, we can attain precision beyond the HS: $F_\omega^\prime=8\delta^2T^2/k^2$ [see Eq. (\red{S65}) in SM] as illustrated in FIG. \ref{fig:f4}(b). 
With $\delta=0.05$, the precision can be $\Delta\omega\approx0.318\omega$ at $4$ ms.
Accordingly, it is feasible to experimentally achieve measurment precisions beyond the HS in a trapped ion with current experimental techniques.

\textit{Conclusion.}---In brief, we take advantage of quantum tricriticality in the aQRM to clarify that quantum criticality is not the sufficient condition for quantum-enhanced metrology.
We also recommend to adiabatically modulate relevant parameters for accomplishments of an exponential super-HS with respect to the evolution time $T$, which results from effective restrains of excitations and critical slowing down. 
Dissipation can contribute to strengthen the super-HS before it is counteracted by decoherence.
Besides, measurment precisions beyond the HS can be reached in an experiment utilizing a trapped ion.
In this tricriticality, we can not explore super-HS with respect to particle number $N$, which may occur in other physical models, such as the anisotropic Dicke model and driven Tavis-Cummings system \cite{baksic2014controlling,zhu2020squeezed}.

\begin{acknowledgments}
	This work was funded National Natural Science Foundation of China (Grants No. 12104363, No. 11974334 and No. 11774332).
	XFZ and ZWZ acknowledge the support from Innovation Program for Quantum Science and Technology 360 (Grant No. 2021ZD0301200).
	YCZ acknowledges the support of National Natural Science Foundation of China (Grant No. 12104359), the National Key R\&D Program of China (2021YFA1401700), Shaanxi Academy of Fundamental Sciences (Mathematics, Physics) (Grant No. 22JSY036), Open project of Low Dimensional Quantum Physics State Key Laboratory (Grant No. KF202104), Xi'an Jiaotong University through the ``Young Top Talents Support Plan" and Basic Research Funding (Grant No. xtr042021012).
\end{acknowledgments}

\bibliography{qm}

	

\end{document}


\supplementarystart

\centerline{\bfseries\large SUPPLEMENTAL MATERIAL}
\vspace{6pt}

\centerline{\bfseries\large Super-Heisenberg scaling in a triple point criticality }
\vspace{6pt}
\centerline{Jia-Ming Cheng, Yong-Chang Zhang, Xiang-Fa Zhou, Zheng-Wei Zhou}


\tableofcontents
\vspace{.5cm}

In this Supplemental Material, we present a number of technical details related to the derivations of our results presented in the main text of the paper.

\section{\label{sec:fi}Quantum Fisher information and excitations near phase boundaries}
We consider a general Hamiltonian of a quantum system undergoing quantum phase transitions, which reads
\begin{align}
	H(\lambda)=H_0+\lambda H_1.
\end{align}
$H_1$ is supposed to be the driving term with controlling parameter $\lambda$. It is assumed that this general Hamiltonian has eigenvalues $E_n(\lambda)$ and corresponding eigenstates $\vert\psi_n(\lambda)\rangle$
\begin{align}
	H(\lambda)\vert\psi_n(\lambda)\rangle=E_n(\lambda)\vert\psi_n(\lambda)\rangle
\end{align}
with $n=0,1,2,\cdots$ and $E_0(\lambda)$ is the ground energy. A quantum phase transition can happen at the critical point $\lambda=\lambda_c$. We further assume that there exists no degeneracy in the ground state, that is $E_0(\lambda)<E_{n\ne0}(\lambda)$ if $\lambda\ne\lambda_c$.
When the controlling parameter varies from $\lambda$ to $\lambda+\delta\lambda$ with $\delta\lambda$ much less than the energy gap, the ground state to first order should be
\begin{align}
	\vert\psi_0(\lambda+\delta\lambda)\rangle=\mathcal{C}\big[ \vert\psi_0(\lambda)\rangle+\delta\lambda\sum_{n\ne0}\frac{H^{n0}_1(\lambda)}{E_0(\lambda)-E_n(\lambda)}\vert\psi_n(\lambda)\rangle \big]
\end{align}
by non-degenerate perturbed theory \cite{sakurai2020modern}, where $\mathcal{C}=\big( 1+\delta\lambda^2\sum_{n\ne0}\vert H^{n0}_1(\lambda)\vert^2/[E_0(\lambda)-E_n(\lambda)]^2 \big)^{-1/2}$ is the normalization constant and $H^{n0}_1(\lambda)=\langle\psi_n(\lambda)\vert H_1\vert\psi_0(\lambda)\rangle$.
Thus, the first order differential of $\vert\psi_0(\lambda)\rangle$ with respect with $\lambda$ should be
\begin{align}
	\vert\partial_\lambda\psi_0(\lambda)\rangle=\lim_{\delta\lambda\to0}\frac{\vert\psi_0(\lambda+\delta\lambda)\rangle-\vert\psi_0(\lambda)\rangle}{\delta\lambda}=\lim_{\delta\lambda\to0}\frac{\mathcal{C}-1}{\delta\lambda}\vert\psi_0(\lambda)\rangle+\sum_{n\ne0}\frac{H^{n0}_1(\lambda)}{E_0(\lambda)-E_n(\lambda)}\vert\psi_n(\lambda)\rangle,
\end{align}
and the QFI relative to $\lambda$ is
\begin{align}
	\label{eq:fl}
	F_\lambda=4[\langle\partial_\lambda\psi_0(\lambda)\vert\partial_\lambda\psi_0(\lambda)\rangle-\vert\langle\partial_\lambda\psi_0(\lambda)\vert\psi_0(\lambda)\rangle\vert^2]
	=4\sum_{n\ne0}\frac{\vert\langle\psi_n(\lambda)\vert H_1\vert\psi_0(\lambda)\rangle\vert^2}{[E_0(\lambda)-E_n(\lambda)]^2}.
\end{align}

Next, we examine excitations during adiabatic evolutions around the critical point. The controlling parameter $\lambda$ now is time-dependent and is changed adiabatically from $\lambda=0$ to a final value $\lambda_f\sim\lambda_c$. At time $t$, its wave-function can be decomposed as in the instantaneous eigen-space
\begin{align}
	\vert\Psi[\lambda(t)]\rangle=\sum_{n}c_n[\lambda(t)]e^{-i\theta_n[\lambda(t)]}\vert\psi_n(\lambda)\rangle
\end{align}
with dynamical phase $\theta_n[\lambda(t)]=\int^t_0E_n(t^\prime)dt^\prime/\hbar $.
We assume the initial state as $c_0[0]=1$ and $c_n[0]=0$ for $n\ne0$. According to the Schr\"{o}dinger equation
\begin{align}
	i\hbar\frac{\partial}{\partial t}\vert\Psi[\lambda(t)]\rangle=H(\lambda)\vert\Psi[\lambda(t)]\rangle,
\end{align}
we can acquire approximately that
\begin{align}
	\frac{\partial}{\partial t}c_m[\lambda(t)]=-\sum_{n}c_n[\lambda(t)]e^{i\big[\theta_m[\lambda(t)]-\theta_n[\lambda(t)]\big]}\langle\psi_m(\lambda)\vert\frac{\partial}{\partial t}\psi_n(\lambda)\rangle.
\end{align}
Based on the time-dependent perturbation theory, we can have that
\begin{align}
	\label{eq:cm}
	c_m[\lambda(t)]=&-\int^t_0e^{i\big[\theta_m[\lambda(t^\prime)]-\theta_0[\lambda(t^\prime)]\big]}\langle\psi_m(t^\prime)\vert\frac{\partial}{\partial t^\prime}\psi_0(t^\prime)\rangle dt^\prime \notag \\
	=&-\int^t_0e^{i\big[\theta_m[\lambda(t^\prime)]-\theta_0[\lambda(t^\prime)]\big]}\dot{\lambda}\frac{\langle\psi_m(t^\prime)\vert H_1\vert\psi_0(t^\prime)\rangle}{E_m(t^\prime)-E_0(t^\prime)} dt^\prime \notag \\
	=&-\int^{\lambda_f}_0e^{i\big[\theta_m[\lambda^\prime]-\theta_0[\lambda^\prime]\big]}\frac{\langle\psi_m(\lambda^\prime)\vert H_1\vert\psi_0(\lambda^\prime)\rangle}{E_m(\lambda^\prime)-E_0(\lambda^\prime)} d\lambda^\prime
\end{align}
for $m\ne0$. Thus the excitation propability is proportional to $\vert\langle\psi_m(\lambda^\prime)\vert H_1\vert\psi_0(\lambda^\prime)\rangle\vert/[E_m(\lambda^\prime)-E_0(\lambda^\prime)]$.

From Eq. (\ref{eq:fl}) and Eq. (\ref{eq:cm}), we can learn that:
\begin{enumerate}
	\item As it is well known that around a normal critical-point of first order or second order, QFI will be divergent because the energy gap closes $E_0(\lambda)-E_n(\lambda)=0$ and meantime it is usually correct that $\langle\psi_n(\lambda)\vert H_1\vert\psi_0(\lambda)\rangle\ne0$, which construct the basis of quantum criticality-enhanced metrology \cite{yang2007ground}. In this criticality, excitations in an adiabatic evolution can no be restrained;
	\item However, around a critical point, such as a triple point, when the two terms $E_0(\lambda)-E_n(\lambda)$ and $\langle\psi_n(\lambda)\vert H_1\vert\psi_0(\lambda)\rangle$ approach zero simultaneously, then QFI will be finite or divergent, which depends on their relative speed of approaching zero. Such a mechanism supplies us with a way to modulate the QFI. This criticality may enable us to devise parameter modulations to restrain excitations in adiabatic evolutions. If the QFI is finite, it would not be useful to quantum metrology.
\end{enumerate}

\section{\label{sec:pd}Phase diagram of the anisotropic quantum Rabi model}
\begin{figure}[htpb]
	\centering
	\includegraphics[width=0.35\textwidth]{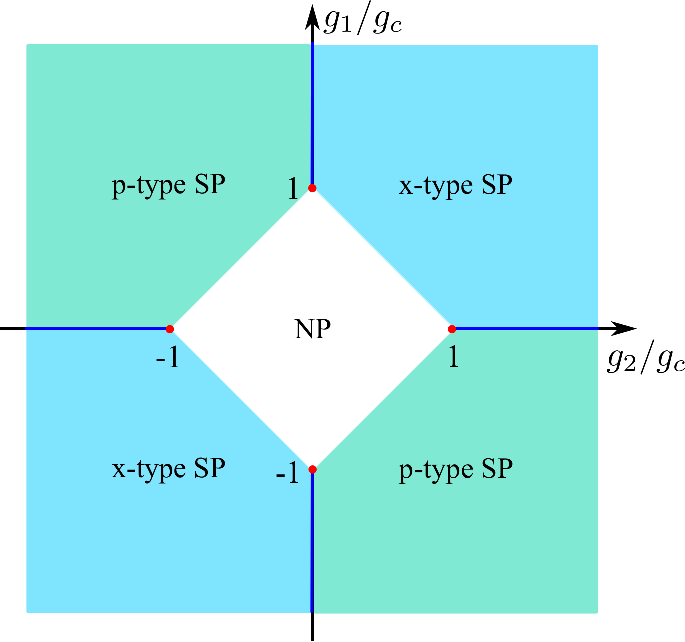}
	\caption{\label{fig:pd}
	Phase diagram of the ground states in plane of $g_1-g_2$. ``NP" and ``SP" are short for normal phase and superradiant phase respectively. The phase transitions from NP to SP are continuous spontaneously breaking the parity symmetry, while those between SPs turn out to be discontinuous.
	There exist four triple points [$(g_1,g_2)=(\pm g_c,0)$, $(0,\pm g_c)$] marked by red dots.
	}
\end{figure}

The anisotropic quantum Rabi model (aQRM) describes interactions between a two-level system and a single-mode light field with different rotating-wave and counter-rotating-wave terms, and can be implemented in diverse physical platforms, such as cavity (circuit) QED systems and trap ions.
Its Hamiltonian can be written as
\begin{align}
	H/\hbar=\frac{\Omega}{2}\sigma_z+\omega a^\dagger a+\frac{g_1}{2}(a^\dagger\sigma_-+a\sigma_+)+\frac{g_2}{2}(a^\dagger\sigma_++a\sigma_-),
\end{align}
where $\sigma_{z,\pm}$ are Pauli operators of the two-level system with transition frequency $\Omega$, ground state $\vert\downarrow\rangle$ and excited state $\vert\uparrow\rangle$, $a^\dagger$ ($a$) is creation (annihilation) operator of the light field with frequency $\omega$, $g_1$ and $g_2$ are their rotating-wave and counter-rotating-wave coupling strength.
It reduces to the quantum Rabi model (QRM) when these two couplings are equal: $g_1=g_2$, and  degrades into the Jaynes-Cummings model or anti-Jaynes-Cummings model if only existing the rotating-wave term ($g_2=0$) or the counter-rotating-wave term ($g_1=0$), which hosts gapless Goldstone model \cite{cheng2023quantum,hwang2016quantum}.
The aQRM Hamiltonian possesses parity symmetry with symmetric operator $\mathcal{P}=\exp[i\pi(a^\dagger a+\sigma_z/2)]$, that is, $[\mathcal{P},H]=0$, because it is easily calculated that
\begin{align}
	\mathcal{P}^\dagger\sigma_\pm\mathcal{P}=-\sigma_\pm,~
	\mathcal{P}^\dagger a\mathcal{P}=-a,~
	\mathcal{P}^\dagger a^\dagger\mathcal{P}=-a^\dagger.
\end{align}
As shown in FIG. \ref{fig:pd}, there exist a normal phase and two superradiant phases in its phase diagram of ground states in the infinite frequency ratio limit $\Omega/\omega\to+\infty$ \cite{liu2017universal}.
Here, for convenience of discussions in the main text, we elaborate characteristics of these phases and nature of corresponding phase transitions.

\textit{Normal phase-}We devote to investigating ground states in the case of $\Omega\gg\omega,g_1,g_2$. When the two interactions $g_1$ and $g_2$ are weak compared with light field frequency $\omega$, its ground state lies in the normal phase. To figure out its concrete form, we first make a unitary Schrieffer-Wolff transformation
\begin{align}
	U_n=\exp[-\xi_n],~
	\xi_n=-\xi^\dagger_n=\frac{g_1}{2(\Omega-\omega)}(a\sigma_+-a^\dagger\sigma_-)+\frac{g_2}{2(\Omega+\omega)}(a^\dagger\sigma_+-a\sigma_-)
\end{align}
and obtain a transformed Hamiltonian $H_n/\hbar=U^\dagger_nHU_n/\hbar$. To second order of $g_1$ and $g_2$, Hamiltonian $H_n$ reads
\begin{align}
	H_n/\hbar=&\frac{\Omega}{2}\sigma_z+\omega a^\dagger a+\frac{1}{8}[ (\frac{g_1^2}{\Omega-\omega}+\frac{g_2^2}{\Omega+\omega})(2a^\dagger a+1)\sigma_z+g_1g_2(\frac{1}{\Omega-\omega}+\frac{1}{\Omega+\omega})(a^{\dagger 2}+a^2)\sigma_z+\frac{g_1^2}{\Omega-\omega}-\frac{g_2^2}{\Omega+\omega} ] \notag \\
	\approx&\frac{\Omega}{2}\sigma_z+\omega a^\dagger a+\frac{g_1^2+g_2^2}{8\Omega}(2a^\dagger a+1)\sigma_z+\frac{g_1g_2}{4\Omega}(a^{\dagger 2}+a^2)\sigma_z+\frac{g_1^2-g_2^2}{8\Omega}
\end{align}
in the infinite frequency ratio limit $\Omega/\omega\to+\infty$.
As the transition frequency $\Omega$ is dominated, this two-level system prefers to lie at its ground state $\vert\downarrow\rangle$. Thus, effective Hamiltonian of the light field becomes
\begin{align}
	H_{np}=\langle\downarrow\vert H_n\vert\downarrow\rangle
	\approx&-\frac{\hbar\Omega}{2}+\hbar\frac{g_1^2-g_2^2}{8\Omega}+\hbar\omega a^\dagger a-\hbar\frac{g_1^2+g_2^2}{8\Omega}(2a^\dagger a+1)-\hbar\frac{g_1g_2}{4\Omega}(a^{\dagger 2}+a^2) \notag \\
	=&-\frac{\hbar\Omega}{2}-\hbar\omega\frac{g_2^2}{g^2_c}+\hbar\omega[ (1-\frac{g_1^2+g_2^2}{g^2_c})a^\dagger a-\frac{g_1g_2}{g^2_c}(a^{\dagger 2}+a^2) ]
\end{align}
with $g_c=2\sqrt{\Omega\omega}$. By applying a squeezing transformation
\begin{align}
	\Gamma(\gamma)=\exp[\frac{\gamma}{2}(a^2-a^{\dagger 2})]~\text{with}~
	e^{2\gamma}=\sqrt{\frac{g^2_c-(g_1+g_2)^2}{g^2_c-(g_1-g_2)^2}},~
	\gamma=\frac{1}{4}\ln\frac{g^2_c-(g_1+g_2)^2}{g^2_c-(g_1-g_2)^2},
\end{align}
the effective Hamiltonian $H_{np}$ can be diagonalized as
\begin{align}
	H^d_{np}=\Gamma^\dagger(\gamma) H_{np}\Gamma(\gamma)=\hbar\Delta a^\dagger a-\frac{\hbar}{2}(\Omega+\omega)+\frac{\hbar\omega}{2}\frac{g_1^2-g^2_2}{g^2_c}+\frac{\hbar\Delta}{2}
\end{align}
with $\Delta=\omega\sqrt{[1-(\frac{g_1-g_2}{g_c})^2][1-(\frac{g_1+g_2}{g_c})^2]}$, where we have used following relationship
\begin{align*}
	\Gamma^\dagger(\gamma) a\Gamma(\gamma)=&a\cosh\gamma-a^\dagger\sinh\gamma=\frac{e^{\gamma}}{2}(a-a^\dagger)+\frac{e^{-\gamma}}{2}(a+a^\dagger).
\end{align*}
It can be found that the energy gap $\Delta$ becomes imaginary when the interactions increase across a critical value so that $\vert g_1-g_2\vert>g_c$ or $\vert g_1+g_2\vert>g_c$, which will lead to instabilities and quantum phase transitions.
At a phase boundary, the energy gap closes, which gives that
\begin{align}
	\vert g_1-g_2\vert=g_c~\text{or}~
	\vert g_1+g_2\vert=g_c.
\end{align}
In the original frame, ground states of the light field should be
\begin{align}
	\vert\psi_{np}\rangle=U_n\Gamma(\gamma)\vert0\rangle=\Gamma(\gamma)\vert0\rangle,
\end{align}
because the unitary Schrieffer-Wolff transformation $U_n=\exp\big[-\sqrt{\frac{\omega}{\Omega}}[\frac{g_1}{g_c}(a\sigma_+-a^\dagger\sigma_-)+\frac{g_2}{g_c}(a^\dagger\sigma_+-a\sigma_-)]\big]=1$ in the limit $\Omega/\omega\to+\infty$.
With these ground states at hand, we can calculate that
\begin{align}
	\langle a\rangle=0,~
	\langle n\rangle=\langle a^\dagger a\rangle=\frac{1}{2}[\cosh(2\gamma)-1],~
	\Delta x=\sqrt{\langle x^2\rangle-\langle x\rangle^2}=\frac{1}{\sqrt{2}}e^{-\gamma},~
	\Delta p=\sqrt{\langle p^2\rangle-\langle p\rangle^2}=\frac{1}{\sqrt{2}}e^{\gamma},
\end{align}
where we have defined position operator $x=(a^\dagger+a)/\sqrt{2}$ and momentum operator $p=i(a^\dagger-a)/\sqrt{2}$.

\textit{Superradiant phase-}When interactions $g_1$ and $g_2$ become strong compared with light field frequency $\omega$, the normal phase turns to be unstable and phase transitions to superradiant states will take place.
To illustrate this phenomenon, we first displace the light field using a displacement transformation $D(\alpha)=\exp(\alpha a^\dagger-\alpha^\ast a)$, and the Hamiltonian is transformed into
\begin{align}
	H^\prime/\hbar=D^\dagger(\alpha)HD(\alpha)/\hbar=\mathcal{H}_q+\omega\vert\alpha\vert^2+\omega(\alpha a^\dagger+\alpha^\ast a)+\omega a^\dagger a+\frac{g_1}{2}(a^\dagger\sigma_-+a\sigma_+)+\frac{g_2}{2}(a^\dagger\sigma_++a\sigma_-)
\end{align}
by using $D^\dagger(\alpha)a D(\alpha)=a+\alpha$. A new Hamiltonian $\mathcal{H}_q$ for the two-level system has form of
\begin{align}
	\mathcal{H}_q=\frac{\Omega}{2}\sigma_z+\frac{1}{2}(g_1\alpha^\ast+g_2\alpha)\sigma_-+\frac{1}{2}(g_1\alpha+g_2\alpha^\ast)\sigma_+=\frac{\Omega}{2}\sigma_z+\frac{1}{2}G(e^{-i\phi}\sigma_-+e^{i\phi}\sigma_+)~\text{with}~
	G=\vert g_1\alpha^\ast+g_2\alpha\vert,
\end{align}
whose eigenvalues are $\epsilon_\pm=\pm\frac{1}{2}\sqrt{\Omega^2+G^2}$. And their corresponding eigenstates can be written as
\begin{align}
	\vert+\rangle=\sin\theta\vert\uparrow\rangle+\cos\theta e^{-i\phi}\vert\downarrow\rangle,~
	\vert-\rangle=\cos\theta e^{i\phi}\vert\uparrow\rangle-\sin\theta\vert\downarrow\rangle,
\end{align}
with $\sin\theta=\frac{1}{\sqrt{2}}\sqrt{1+\frac{\Omega}{\sqrt{\Omega^2+G^2}}}$, $\cos\theta=\frac{1}{\sqrt{2}}\sqrt{1-\frac{\Omega}{\sqrt{\Omega^2+G^2}}}$, from which we can get that
\begin{align}
	&\vert\uparrow\rangle=\sin\theta\vert+\rangle+\cos\theta e^{-i\phi}\vert-\rangle,~
	\vert\downarrow\rangle=\cos\theta e^{i\phi}\vert+\rangle-\sin\theta\vert-\rangle,\\
	&\sigma_+=\sigma_-^\dagger=\vert\uparrow\rangle\langle\downarrow\vert=\frac{1}{2}\sin(2\theta)e^{-i\phi}\tau_z-\sin^2\theta\tau_++\cos^2\theta e^{-i2\phi}\tau_-.
\end{align}
Here we have defined new Pauli operators in the eigen-space
\begin{align}
	\tau_z=\vert+\rangle\langle+\vert-\vert-\rangle\langle-\vert,~
	\tau_+=\tau^\dagger_-=\vert+\rangle\langle-\vert.
\end{align}
In this basis, the transformed Hamiltonian turns out to be
\begin{align}
	H^\prime/\hbar=&\omega(\alpha a^\dagger+\alpha^\ast a)+\frac{1}{4}\sin(2\theta)\tau_z[(g_1e^{-i\phi}+g_2e^{i\phi})a+(g_1e^{i\phi}+g_2e^{-i\phi})a^\dagger] \notag \\
	&+\frac{1}{2}\sqrt{\Omega^2+G^2}\tau_z+\omega\vert\alpha\vert^2+\omega a^\dagger a-\frac{1}{2}\sin^2\theta[g_1(a\tau_++a^\dagger\tau_-)+g_2(a^\dagger\tau_++a\tau_-)] \notag \\
	&+\frac{1}{2}\cos^2\theta[g_1(e^{i2\phi}a^\dagger\tau_++e^{-i2\phi}a\tau_-)+g_2(e^{-i2\phi}a^\dagger\tau_-+e^{i2\phi}a\tau_+)].
\end{align}
As transition energy of the two-level system is dominated, the low-energy physics is constrained in the subspace of $\vert-\rangle$. In addition, the parity symmetry $\mathcal{P}^\prime=\exp[i\pi(a^\dagger a+\tau_z/2)]$ should be satisfied, which demands that
\begin{align}
	\omega\alpha-\frac{1}{4}\sin(2\theta)(g_1e^{i\phi}+g_2e^{-i\phi})=0.
\end{align}
Solving this equation, we can know that
\begin{align}
	\begin{cases}
		\alpha=\pm i\frac{\Omega}{\vert g_1-g_2\vert}\sqrt{(\frac{g_1-g_2}{g_c})^4-1},~G=\vert(g_1-g_2)\alpha\vert \\
		e^{i2\phi}=e^{-i2\phi}=-1,~\sin^2\theta=\frac{1}{2}[1+(\frac{g_c}{g_1-g_2})^2],
	\end{cases}
\end{align}
or
\begin{align}
	\begin{cases}
		\alpha=\pm \frac{\Omega}{\vert g_1+g_2\vert}\sqrt{(\frac{g_1+g_2}{g_c})^4-1},~G=\vert(g_1+g_2)\alpha\vert \\
		e^{i2\phi}=e^{-i2\phi}=1,~\sin^2\theta=\frac{1}{2}[1+(\frac{g_c}{g_1+g_2})^2].
	\end{cases}
\end{align}
Thus, when the displacement $\alpha$ is real, the transformed Hamiltonian becomes
\begin{align}
	H^\prime/\hbar=&\frac{\Omega}{4}[(\frac{g_1+g_2}{g_c})^2-(\frac{g_c}{g_1+g_2})^2]+\frac{\Omega^\prime}{2}\tau_z+\omega a^\dagger a+\frac{g_1^\prime}{2}(a\tau_++a^\dagger\tau_-)+\frac{g_2^\prime}{2}(a^\dagger\tau_++a\tau_-)
\end{align}
with $\Omega^\prime=\Omega(\frac{g_1+g_2}{g_c})^2$, $g_1^\prime=-\frac{1}{2}(g_1-g_2+\frac{g^2_c}{g_1+g_2})$ and $g_2^\prime=\frac{1}{2}(g_1-g_2-\frac{g^2_c}{g_1+g_2})$;
when the displacement $\alpha$ is pure complex, the transformed Hamiltonian becomes
\begin{align}
	H^\prime/\hbar=&\frac{\Omega}{4}[(\frac{g_1-g_2}{g_c})^2-(\frac{g_c}{g_1-g_2})^2]+\frac{\Omega^\prime}{2}\tau_z+\omega a^\dagger a+\frac{g_1^\prime}{2}(a\tau_++a^\dagger\tau_-)+\frac{g_2^\prime}{2}(a^\dagger\tau_++a\tau_-)
\end{align}
with $\Omega^\prime=\Omega(\frac{g_1-g_2}{g_c})^2$, $g_1^\prime=-\frac{1}{2}(g_1+g_2+\frac{g^2_c}{g_1-g_2})$ and $g_2^\prime=-\frac{1}{2}(g_1+g_2-\frac{g^2_c}{g_1-g_2})$.
The Hamiltonian $H^\prime$ after making a displacement transformation has the same form as the original Hamiltonian $H$ except for the renormalized transition frequency $\Omega^\prime$, interactions $g^\prime_1$ and $g^\prime_2$.
We can diagonalize it using similar Schrieffer-Wolff transformations and squeezing transformations.
Specifically, we make a unitary Schrieffer-Wolff transformation
\begin{align}
	U_s=\exp[-\xi_s],~
	\xi_s=-\xi^\dagger_s=\frac{g^\prime_1}{2(\Omega^\prime-\omega)}(a\tau_+-a^\dagger\tau_-)+\frac{g^\prime_2}{2(\Omega^\prime+\omega)}(a^\dagger\tau_+-a\tau_-)
\end{align}
and obtain a transformed Hamiltonian $H_s/\hbar=U^\dagger_sH^\prime U_s/\hbar$. To second order of $g_1^\prime$ and $g_2^\prime$, Hamiltonian $H_s$ reads
\begin{align}
	H_s/\hbar=&\frac{\Omega^\prime}{2}\tau_z+\omega a^\dagger a+\frac{1}{8}[ (\frac{g_1^{\prime2}}{\Omega^\prime-\omega}+\frac{g_2^{\prime2}}{\Omega^\prime+\omega})(2a^\dagger a+1)\tau_z+g^\prime_1g^\prime_2(\frac{1}{\Omega^\prime-\omega}+\frac{1}{\Omega^\prime+\omega})(a^{\dagger 2}+a^2)\tau_z+\frac{g_1^{\prime2}}{\Omega^\prime-\omega}-\frac{g_2^{\prime2}}{\Omega^\prime+\omega} ] \notag \\
	=&\frac{\Omega^\prime}{2}\tau_z+\omega a^\dagger a+\frac{g_1^{\prime2}+g_2^{\prime2}}{8\Omega^\prime}(2a^\dagger a+1)\tau_z+\frac{g_1^\prime g^\prime_2}{4\Omega^\prime}(a^{\dagger 2}+a^2)\tau_z+\frac{g_1^{\prime2}-g_2^{\prime2}}{8\Omega^\prime}
\end{align}
in the limit $\Omega/\omega\to+\infty$, where we have ignored a constant term $\frac{\Omega}{4}[(\frac{g_1\pm g_2}{g_c})^2-(\frac{g_c}{g_1\pm g_2})^2]$.
As the two-level system prefers to stay at its ground state $\vert-\rangle$, effective Hamiltonian of the light field becomes
\begin{align}
	H_{sp}=\langle-\vert H_s\vert-\rangle
	=&-\frac{\hbar\Omega^\prime}{2}+\hbar\frac{g_1^{\prime2}-g_2^{\prime2}}{8\Omega^\prime}+\hbar\omega a^\dagger a-\hbar\frac{g_1^{\prime2}+g_2^{\prime2}}{8\Omega^\prime}(2a^\dagger a+1)-\hbar\frac{g_1^\prime g_2^\prime}{4\Omega^\prime}(a^{\dagger 2}+a^2) \notag \\
	=&-\frac{\hbar\Omega^\prime}{2}-\hbar\omega\frac{g_2^{\prime2}}{g^{\prime2}_c}+\hbar\omega[ (1-\frac{g_1^{\prime2}+g_2^{\prime2}}{g^{\prime2}_c})a^\dagger a-\frac{g_1^\prime g_2^\prime}{g^{\prime2}_c}(a^{\dagger 2}+a^2) ]
\end{align}
with $g^\prime_c=2\sqrt{\Omega^\prime\omega}$.
Then applying a squeezing transformation
\begin{align}
	\Gamma(\gamma^\prime)=\exp[\frac{\gamma^\prime}{2}(a^2-a^{\dagger 2})]~\text{with}~
	e^{2\gamma^\prime}=\sqrt{\frac{g^{\prime2}_c-(g_1^\prime+g_2^\prime)^2}{g^{\prime2}_c-(g_1^\prime-g_2^\prime)^2}},~
	\gamma^\prime=\frac{1}{4}\ln\frac{g^{\prime2}_c-(g_1^\prime+g_2^\prime)^2}{g^{\prime2}_c-(g_1^\prime-g_2^\prime)^2},
\end{align}
the effective Hamiltonian $H_{sp}$ can be diagonalized as
\begin{align}
	H^d_{sp}=\Gamma^\dagger(\gamma^\prime) H_{sp}\Gamma(\gamma^\prime)=\hbar\Delta^\prime a^\dagger a-\frac{\hbar}{2}(\Omega^\prime+\omega)+\frac{\hbar\omega}{2}\frac{g_1^{\prime2}-g^{\prime2}_2}{g^{\prime2}_c}+\frac{\hbar\Delta^\prime}{2}
\end{align}
with $\Delta^\prime=\omega\sqrt{[1-(\frac{g_1^\prime-g_2^\prime}{g^\prime_c})^2][1-(\frac{g_1^\prime+g_2^\prime}{g^\prime_c})^2]}$.
So, if the displacement $\alpha$ is real, we can obtain the squeezing factor
\begin{align}
	\gamma^\prime=\frac{1}{4}\ln\frac{1-(\frac{g_c}{g_1+g_2})^4}{1-(\frac{g_1-g_2}{g_1+g_2})^2}=\frac{1}{4}\big(\ln[1-(\frac{g_c}{g_1+g_2})^4]-\ln[1-(\frac{g_1-g_2}{g_1+g_2})^2]\big)
\end{align}
and diagonalized Hamiltonian
\begin{align}
	H^d_{sp}=\hbar\Delta^\prime a^\dagger a-\frac{\hbar\Omega}{4}[(\frac{g_1+g_2}{g_c})^2+(\frac{g_c}{g_1+g_2})^2]-\frac{\hbar\omega}{2}+\frac{\hbar\omega}{2}\frac{g_1-g_2}{g_1+g_2}(\frac{g_c}{g_1+g_2})^2+\frac{\hbar\Delta^\prime}{2}
\end{align}
with $\Delta^\prime=\omega\sqrt{[1-(\frac{g_1-g_2}{g_1+g_2})^2][1-(\frac{g_c}{g_1+g_2})^4]}$;
if the displacement $\alpha$ is pure complex, it can be known that the squeezing factor should be
\begin{align}
	\gamma^\prime=\frac{1}{4}\ln\frac{1-(\frac{g_1+g_2}{g_1-g_2})^2}{1-(\frac{g_c}{g_1-g_2})^4}=\frac{1}{4}\big(\ln[1-(\frac{g_1+g_2}{g_1-g_2})^2]-\ln[1-(\frac{g_c}{g_1-g_2})^4]\big).
\end{align}
Then the diagonalized Hamiltonian becomes
\begin{align}
	H^d_{sp}=\hbar\Delta^\prime a^\dagger a-\frac{\hbar\Omega}{4}[(\frac{g_1-g_2}{g_c})^2+(\frac{g_c}{g_1-g_2})^2]-\frac{\hbar\omega}{2}+\frac{\hbar\omega}{2}\frac{g_1+g_2}{g_1-g_2}(\frac{g_c}{g_1-g_2})^2+\frac{\hbar\Delta^\prime}{2}
\end{align}
with $\Delta^\prime=\omega\sqrt{[1-(\frac{g_1+g_2}{g_1-g_2})^2][1-(\frac{g_c}{g_1-g_2})^4]}$.
In the original frame, ground states of the light field are
\begin{align}
	\vert\psi_{sp}\rangle=D(\alpha)U_s\Gamma(\gamma^\prime)\vert0\rangle=D(\alpha)\Gamma(\gamma^\prime)\vert0\rangle
\end{align}
in the limit $\Omega/\omega\to+\infty$.
With these ground states in mind, we can acquire that
\begin{align}
	&\langle a\rangle=\alpha,~
	\langle a^\dagger a\rangle=\vert\alpha\vert^2+\frac{1}{2}[\cosh(2\gamma^\prime)-1], ~
	\langle x\rangle=\frac{1}{\sqrt{2}}(\alpha^{\ast}+\alpha),~
	\langle p\rangle=\frac{i}{\sqrt{2}}(\alpha^{\ast}-\alpha),~
	\Delta x=\frac{e^{-\gamma^\prime}}{\sqrt{2}},~
	\Delta p=\frac{e^{\gamma^\prime}}{\sqrt{2}}.
\end{align}
Whether the displacement $\alpha$ is real or not depends on their corresponding ground energies. When they are equal, we can have that
\begin{align}
	g_1=0~ \text{or}~g_2=0.
\end{align}

\begin{figure}[htbp]
	\centering
	\includegraphics[width=0.43\textwidth]{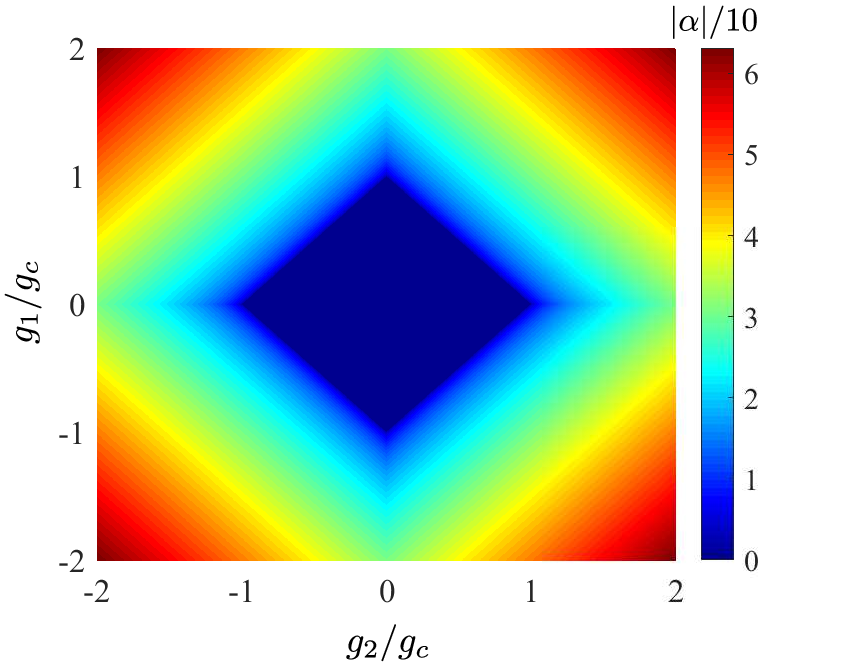}
	\caption{\label{fig:alpha}
	Amplitude of the displacement $\alpha$ in case of $\omega=1$, $\Omega=10^3$.
	}
\end{figure}
\begin{figure}[htbp]
	\centering
	\includegraphics[width=0.9\textwidth]{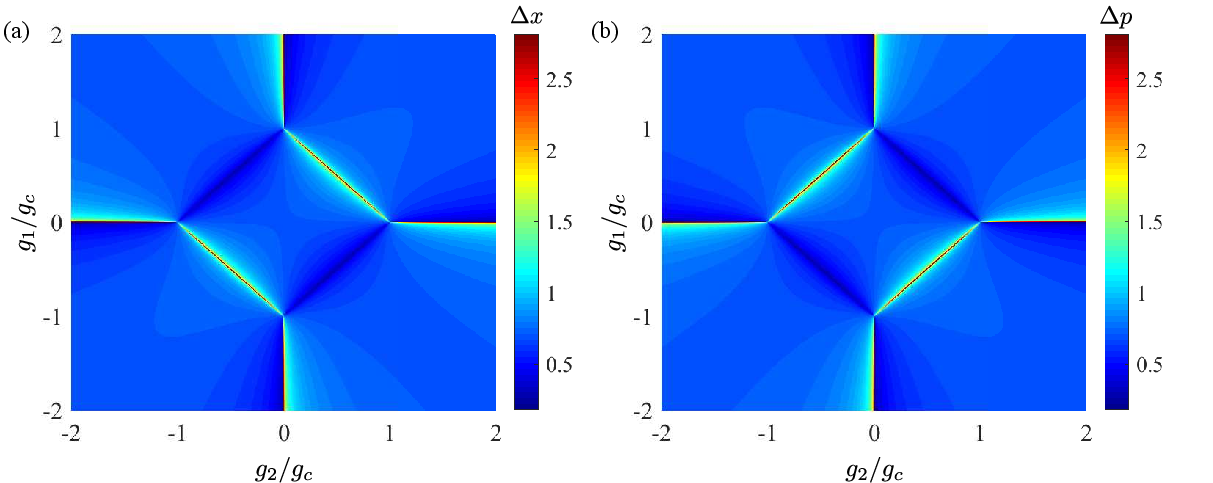}
	\caption{\label{fig:deltaxp}
	Standard deviation $\Delta x$ (a) and $\Delta p$ (b) in the $g_1-g_2$ plane.
	}
\end{figure}

Therefore, when $\vert g_1+g_2\vert>g_c$ and $g_1g_2>0$, $\alpha$ is real, which gives that $\langle x\rangle\ne0$ and $\langle p\rangle=0$. The SP is named as x-type superradiant phase; when $\vert g_1-g_2\vert>g_c$ and $g_1g_2<0$, $\alpha$ is pure complex, which leads to that $\langle x\rangle=0$ and $\langle p\rangle\ne0$. This SP is of p-type.
There exist two kinds of superradiant phases in its phase diagram and the phase transition between them is of first-order, in addition, phase transitions from normal phase to superradiant phases are second-order [see FIG. \ref{fig:alpha}].
From FIG. \ref{fig:deltaxp}, we can learn that squeezing of $x$ or $p$ is maximum at these phase boundaries far from the four triple points ($g_1=0$, $g_2=\pm g_c$, or $g_2=0$, $g_1=\pm g_c$), where $\Delta x=\Delta p=1/\sqrt{2}$ and no squeezing exists.
It also indicates that wave functions close to phase boundaries $\vert g_1\pm g_1\vert=g_c$ change more drastically than that at the four triple points.

\section{\label{sec:app}Approximately adiabatic evolution in the normal phase}
In order to accomplish a criticality-enhanced parameter estimation, we first prepare the aQRM in its ground states and then adiabatically bring the system in close proximity to a phase boundary.
In particular, we show an example of measuring the light-field frequency $\omega$ and find analytical expressions of quantum Fisher information in the normal phase side by modulating interactions $g_1$ and $g_2$ slowly enough to ensure approximate adiabaticity.
So the time-dependent Hamiltonian for adiabatic evolution is
\begin{align}
	H_{np}(t)=-\frac{\hbar\Omega}{2}-\hbar\omega\frac{g_2^2(t)}{g^2_c}+\hbar\omega[ (1-\frac{g_1^2(t)+g_2^2(t)}{g^2_c})a^\dagger a-\frac{g_1(t)g_2(t)}{g^2_c}(a^{\dagger 2}+a^2) ].
\end{align}
Its instantaneous eigenstates $\vert\psi_n(t)\rangle$ are given by squeezed Fock states
\begin{align}
	H_{np}(t)\vert\psi_n(t)\rangle=\hbar\omega_n(t)\vert\psi_n(t)\rangle,~
	\vert\psi_n(t)\rangle=\Gamma[\gamma(t)]\vert n\rangle=e^{\frac{\gamma(t)}{2}(a^2-a^{\dagger2})}\vert n\rangle
\end{align}
with integer $n=0,1,2,\cdots$, eigen-energy $\hbar\omega_n(t)=E_0(t)+n\hbar\Delta(t)$, ground energy $E_0(t)=-\frac{\hbar}{2}(\Omega+\omega)+\frac{\hbar\omega}{2}\frac{g_1^2(t)-g^2_2(t)}{g^2_c}+\frac{\hbar}{2}\Delta(t)$, energy gap $\Delta(t)=\omega\sqrt{[1-(\frac{g_1(t)-g_2(t)}{g_c})^2][1-(\frac{g_1(t)+g_2(t)}{g_c})^2]}$ and squeezing factor $\gamma(t)=\frac{1}{4}\ln\frac{g^2_c-[g_1(t)+g_2(t)]^2}{g^2_c-[g_1(t)-g_2(t)]^2}$.
If the evolution is adiabatic, states of the light-field at time $t$ should be its instantaneous eigenstate $\vert\psi_0(t)\rangle=\vert\psi_{np}\rangle$.
However, it is inevitable to excite the light-field in an actual state evolving, so we intend to find out suitable slow ramp rates needed to approximately follow the instantaneous ground state $\vert\psi_{np}\rangle$. 
At time $t$, its wave-function can be decomposed as in this instantaneous eigen-space
\begin{align}
	\vert\psi(t)\rangle=\sum_{n}c_n(t)e^{-i\theta_n(t)}\vert\psi_n(t)\rangle
\end{align}
with dynamical phase $\theta_n(t)=\int^t_0\omega_n(t^\prime)dt^\prime$.
So the initial conditions are $c_0(0)=1$ and $c_n(0)=0$ for $n\ne0$. We can compute the evolution using Schr\"{o}dinger equation
\begin{align}
	i\hbar\frac{\partial}{\partial t}\psi(t)=H_{np}(t)\psi(t),
\end{align}
which leads to
\begin{align}
	\frac{\partial}{\partial t}c_m(t)=-\sum_{n}c_n(t)e^{i[\theta_m(t)-\theta_n(t)]}\langle\psi_m(t)\vert\frac{\partial}{\partial t}\psi_n(t)\rangle.
\end{align}
According to time-dependent perturbation theory, it gives that
\begin{align}
	c_m(t)=&-\int^t_0e^{i[\theta_m(t^\prime)-\theta_0(t^\prime)]}\langle\psi_m(t^\prime)\vert\frac{\partial}{\partial t^\prime}\psi_0(t^\prime)\rangle dt^\prime \notag \\
	=&-\int^t_0e^{i[\theta_m(t^\prime)-\theta_0(t^\prime)]}\langle m\vert e^{\frac{\gamma(t^\prime)}{2}(a^{\dagger2}-a^2)}\frac{1}{2}(\frac{\partial}{\partial t^\prime}\gamma(t^\prime))(a^2-a^{\dagger2})e^{\frac{\gamma(t^\prime)}{2}(a^{2}-a^{\dagger2})}\vert0\rangle dt^\prime \notag \\
	=&\int^t_0e^{i[\theta_m(t^\prime)-\theta_0(t^\prime)]}\frac{1}{\sqrt{2}}[\frac{\partial}{\partial t^\prime}\gamma(t^\prime)]\delta_{m,2} dt^\prime,
\end{align}
from which we can know that only transitions to the second-excited state should be taken into account, and we can rewrite that
\begin{align}
	c_2(t)=&\frac{1}{\sqrt{2}}\int^t_0e^{i[\theta_2(t^\prime)-\theta_0(t^\prime)]}[\frac{\partial}{\partial t^\prime}\gamma(t^\prime)] dt^\prime \notag \\
	=&\frac{1}{\sqrt{2}}\int^t_0e^{i[\theta_2(t^\prime)-\theta_0(t^\prime)]}(-\frac{1}{2})\Big( \frac{[g_1(t^\prime)+g_2(t^\prime)][\dot{g}_1(t^\prime)+\dot{g}_2(t^\prime)]}{g_c^2-[g_1(t^\prime)+g_2(t^\prime)]^2}-\frac{[g_1(t^\prime)-g_2(t^\prime)][\dot{g}_1(t^\prime)-\dot{g}_2(t^\prime)]}{g_c^2-[g_1(t^\prime)-g_2(t^\prime)]^2} \Big) dt^\prime .
\end{align}
So for approximate adiabaticity,  it is demanded that
\begin{align}
	\vert c_2(t)\vert^2\ll1,
\end{align}
which means that there are almost no transitions form ground states to excitations.

\section{Quantum Fisher information close to the triple points}
The precision of measuring light-field frequency $\omega$ is bounded by quantum Cram\'er-Rao bound: $\Delta\omega\ge1/\sqrt{\nu F_\omega}$ \cite{escher2011quantum}, where $\nu$ is the number of independent measurements and $F_\omega$ is the quantum Fisher information (QFI) relative to an interested parameter $\omega$.
Since the final point of our adiabatic evolution is near a phase boundary and the final state is a normal state, the QFI can be computed exactly as $F_\omega=4[ \langle\partial_\omega\psi_{np}\vert\partial_\omega\psi_{np}\rangle-\vert\langle\partial_\omega\psi_{np}\vert\psi_{np}\rangle\vert^2 ]=2(\frac{\partial\gamma}{\partial\omega})^2$, whose concrete form is
\begin{align}
	F_\omega=\frac{g_c^4}{8\omega^2}[\frac{1}{g^2_c-(g_1+g_2)^2}-\frac{1}{g_c^2-(g_1-g_2)^2}]^2=\frac{2g_c^4}{\omega^2}\frac{g^2_1g^2_2}{[(g_c+g_1)^2-g_2^2]^2(g_c-g_1+g_2)^2(g_c-g_1-g_2)^2}.
\end{align}
It clearly manifests characteristics of the QFI $F_\omega$ at a critical point:
\begin{enumerate}
	\item Near a phase boundary $\vert g_1+g_2\vert=g_c$ or $\vert g_1-g_2\vert=g_c$, the QFI is divergent. However, near a triple point,  whether it is divergent depends on the relations between $g_1$ and $g_2$;
	\item For example, near the triple point $(g_1,g_2)=(g_c,0)$, we assume $1-\frac{g_1}{g_c}=k(\frac{g_2}{g_c})^\beta (k>1, 0<\beta\le1)$. If $0<\beta<\frac{1}{2}$, QFI $F_\omega\approx \frac{1}{8\omega^2k^4}(\frac{g_2}{g_c})^{2(1-2\beta)}\to0$; if $\beta=\frac{1}{2}$, QFI $F_\omega\approx\frac{1}{8\omega^2k^4}$ is finite; when $\frac{1}{2}<\beta<1$, QFI $F_\omega\approx\frac{1}{8\omega^2k^4}(\frac{g_2}{g_c})^{2(1-2\beta)}\to\infty$ is divergent;  when $\beta=1$, QFI $F_\omega\approx\frac{1}{8\omega^2(k^2-1)^2}(\frac{g_2}{g_c})^{-2}\to\infty$ is divergent. Moreover, when $g_2=0$, QFI $F_\omega=0$.
\end{enumerate}
When QFI is not divergent near a critical point, this criticality will not to be beneficial to quantum-enhanced metrology. On the other hand, when QFI is divergent, adiabatic evolutions are usually needed in criticality-enhanced metrology.
However, the adiabatic evolution time $T$ will also lengthen to infinity accompanying with closed energy gaps, which is known as critical slowing down.
So in order to assess the performances of this adiabatic protocol, it is necessary to consider the evolution time $T$ and average photon number $N$ that are used in the critical sensing.

\subsection{\label{sec:sup}Super-Heisenberg scaling in adiabatic modulations along a straight line}
To inspect QFI close to the triple point $(g_1,g_2)=(g_c,0)$, we adapt adiabatically interactions $g_1$ and $g_2$ along the line using a general adaptive manner
\begin{align}
	\frac{g_1(t)}{g_c}=1-k\frac{g_2(t)}{g_c},~ k>1,~
	\frac{g_2(t)}{g_c}=1/k-\int^t_0 v(t^\prime)dt^\prime,
\end{align}
from which it can be obtained that
\begin{align*}
	g_1+g_2=g_c+(1-k)g_2,~
	g_1-g_2=g_c-(1+k)g_2,~
	\frac{\dot{g}_2(t)}{g_c}=-v(t),~\frac{\dot{g}_1(t)}{g_c}=k v(t).
\end{align*}
Then, the coefficient of excitations in evolving wave-function becomes
\begin{align}
	\label{eq:c2}
	c_2(t)=&\frac{1}{\sqrt{2}}\int^t_0e^{i[\theta_2(t^\prime)-\theta_0(t^\prime)]}(-\frac{1}{2})\Big( \frac{[g_1(t^\prime)+g_2(t^\prime)][\dot{g}_1(t^\prime)+\dot{g}_2(t^\prime)]}{g_c^2-[g_1(t^\prime)+g_2(t^\prime)]^2}-\frac{[g_1(t^\prime)-g_2(t^\prime)][\dot{g}_1(t^\prime)-\dot{g}_2(t^\prime)]}{g_c^2-[g_1(t^\prime)-g_2(t^\prime)]^2} \Big) dt^\prime  \notag \\
	=&\frac{\sqrt{2}}{4}\int^t_0e^{i[\theta_2(t^\prime)-\theta_0(t^\prime)]}\Big( \frac{[g_1(t^\prime)+g_2(t^\prime)][\partial_{g_2}g_1(t^\prime)+1]}{g_c^2-[g_1(t^\prime)+g_2(t^\prime)]^2}-\frac{[g_1(t^\prime)-g_2(t^\prime)][\partial_{g_2}g_1(t^\prime)-1]}{g_c^2-[g_1(t^\prime)-g_2(t^\prime)]^2} \Big)(-)g_cv(t^\prime) dt^\prime \notag \\
	=&\frac{\sqrt{2}}{4}\int^g_{g_c/k}e^{i[\theta_2(g_2)-\theta_0(g_2)]}\Big( \frac{(g_1+g_2)(\partial_{g_2}g_1+1)}{g_c^2-(g_1+g_2)^2}-\frac{(g_1-g_2)(\partial_{g_2}g_1-1)}{g_c^2-(g_1-g_2)^2} \Big) dg_2 \notag \\
	=&\frac{1}{\sqrt{2}}\int^{g}_{g_c/k}e^{i\Theta(g_2)}\mathcal{F}(g_2)dg_2,
\end{align}
where we have defined two functions $\mathcal{F}(g_2)$ and $\Theta(g_2)$ 
\begin{align}
	\mathcal{F}(g_2)=&\frac{1}{2}\Big( \frac{(g_1+g_2)(\partial_{g_2}g_1+1)}{g_c^2-(g_1+g_2)^2}-\frac{(g_1-g_2)(\partial_{g_2}g_1-1)}{g_c^2-(g_1-g_2)^2} \Big) \notag \\
	=&\frac{-g_c}{4g_c^2-4k g_cg_2+(k^2-1)g_2^2}\approx-\frac{1}{4g_c} \notag \\
	\Theta(g_2)=&\theta_2(g_2)-\theta_0(g_2)=-\int^{g_2}_{g_c/k}\frac{2\Delta(g^\prime)}{g_cv(g^\prime)}dg^\prime.
\end{align}
It should be noted that function $\mathcal{F}(g_2)=-\partial_{g_2}\gamma$ is finite, but not divergent at the triple point $(g_1,g_1)=(g_c,0)$, which is quite different from that in the QRM \cite{garbe2020critical},
and can weaken the well-known critical slowing down effect to a certain degree so that its optimal ramp rates $v(g_2)$ can be much greater than that in criticality metrology utilizing the QRM \cite{garbe2020critical}.
In other words, finite function $\mathcal{F}(g_2)$ can greatly decrease the adiabatic evolution time $T$, which is a fascinating and valuable characteristic for actual quantum metrology with limited coherence time.
To ensure that $c_2(t)$ remains small during the evolution, the ramp rate should be small, that is $v(g_2)\ll1$, so that $\Theta(g_2)$ is large and the exponential of the integral in Eq. (\ref{eq:c2}) oscillates fast, canceling the integral.
The exponential term should oscillate much faster than evolution of $\mathcal{F}(g_2)$, so we need to have that
\begin{align}
	\vert\frac{\partial_{g_2}\mathcal{F}(g_2)}{\mathcal{F}(g_2)}\vert\ll\vert\partial_{g_2}\Theta(g_2)\vert=2\frac{\Delta(g_2)}{\vert v(g_2)\vert}.
\end{align}
Around this triple point, as the energy gap can be approximated as
\begin{align}
	\Delta(g_2)=&\omega\sqrt{\big(1-[1-(1+k)\frac{g_2}{g_c}]^2\big)\big(1-[1-(k-1)\frac{g_2}{g_c}]^2\big)}  \notag \\
	=&\omega\sqrt{(k^2-1)[2-(k+1)\frac{g_2}{g_c}][2-(k-1)\frac{g_2}{g_c}]}\frac{g_2}{g_c}
	\approx2\omega(k^2-1)^{1/2}\frac{g_2}{g_c},
\end{align}
the optimal ramp rate should be set as
\begin{align}
	v(g_2)=\frac{2\delta}{k}\Delta(g_2)\approx\frac{4\delta\omega}{k}\sqrt{k^2-1}\frac{g_2}{g_c}
\end{align}
with a small parameter $\delta\ll1$, which decreases to zero in the same way with that of energy gap. The excitation probability can be approximated as $\vert c_2(g_2)\vert^2\approx\frac{\delta^2}{32k^2}\ll1$.
The average photon number $N$ and evolution time $T$ can be calculated as
\begin{align}
	N=&\frac{1}{4}[\sqrt{\frac{g^2_c-(g_1+g_2)^2}{g^2_c-(g_1-g_2)^2}}+\sqrt{\frac{g^2_c-(g_1-g_2)^2}{g^2_c-(g_1+g_2)^2}}]-\frac{1}{2}
	\approx\frac{1}{4}(\sqrt{\frac{k-1}{k+1}}+\sqrt{\frac{k+1}{k-1}})-\frac{1}{2}, \\
	T=&\int^{g_2}_{g_c/k}\frac{1}{-g_cv(g_2^\prime)}dg_2^\prime\approx -\frac{1}{4\delta\omega}(1-\frac{1}{k^2})^{-1/2}\ln(\frac{g_2}{g_c}). 
\end{align}
As QFI around the triple point $(g_1,g_2)=(g_c,0)$ is
\begin{align}
	F_\omega=\frac{1}{8\omega^2}\big( [1-(\frac{g_1+g_2}{g_c})^2]^{-1}-[1-(\frac{g_1-g_2}{g_c})^2]^{-1} \big)^2
	\approx\frac{1}{8\omega^2}\frac{1}{(k^2-1)^2}(\frac{g_2}{g_c})^{-2},
\end{align}
it can be obtained that
\begin{align}
	F_\omega\approx&\frac{1}{8\omega^2}\frac{1}{(k^2-1)^2}e^{8\delta\omega\sqrt{1-\frac{1}{k^2}}T},
\end{align}
from which we can know that the sensing protocol proposed around the triple points can greatly surpass the Heisenberg scaling with respect to the adiabatic evolution time $T$.
This exponential scaling can make it possible to overcome the dilemma of finite coherence time in actual critical-metrology.

If we set a slower ramp rate
\begin{align}
	v(g_2)=\frac{2\delta}{k\omega}\Delta^2(g_2)\approx\frac{8\delta\omega}{k}(k^2-1)(\frac{g_2}{g_c})^2,
\end{align}
the evolution time $T$ can be calculated as
\begin{align}
	T=\int^{g_2}_{g_c/k}\frac{1}{-g_cv(g_2^\prime)}dg_2^\prime\approx \frac{k}{8\delta}\frac{1}{\omega(k^2-1)}(\frac{g_2}{g_c})^{-1}. 
\end{align}
Thus we can obtain the Heisenberg scaling
\begin{align}
	F_\omega\approx\frac{1}{8\omega^2}\frac{1}{(k^2-1)^2}(\frac{g_2}{g_c})^{-2}\approx8\delta^2T^2/k^2.
\end{align}

\subsection{\label{sec:suphei}Super-Heisenberg scaling in adiabatic modulations along a parabola}
In this section, we study QFI close to the triple point $(g_1,g_2)=(g_c,0)$ by adapting adiabatically interactions $g_1$ and $g_2$ along the parabola in a general adaptive manner
\begin{align}
	\frac{g_1(t)}{g_c}=[1-k\frac{g_2(t)}{g_c}]^2,~ k>1,~
	\frac{g_2(t)}{g_c}=\frac{1}{k}-\int^t_0 v(t^\prime)dt^\prime.
\end{align}
It can be easily obtained that
\begin{align*}
	&g_c+g_1\pm g_2=g_c[2-(2k\mp1)\frac{g_2}{g_c}+k^2(\frac{g_2}{g_c})^2],~ \\
	&g_c-g_1\pm g_2=g_c[(2k\pm1)-k^2\frac{g_2}{g_c}]\frac{g_2}{g_c},~ \\
	&\partial_{g_2}g_1=2k(k\frac{g_2}{g_c}-1),~\dot{g}_2(t)=-g_cv(t).
\end{align*}
Then, the coefficient of excitations in evolving wave-function becomes
\begin{align}
	c_2(t)=&\frac{1}{\sqrt{2}}\int^t_0e^{i[\theta_2(t^\prime)-\theta_0(t^\prime)]}(-\frac{1}{2})\Big( \frac{[g_1(t^\prime)+g_2(t^\prime)][\dot{g}_1(t^\prime)+\dot{g}_2(t^\prime)]}{g_c^2-[g_1(t^\prime)+g_2(t^\prime)]^2}-\frac{[g_1(t^\prime)-g_2(t^\prime)][\dot{g}_1(t^\prime)-\dot{g}_2(t^\prime)]}{g_c^2-[g_1(t^\prime)-g_2(t^\prime)]^2} \Big) dt^\prime  \notag \\
	=&\frac{\sqrt{2}}{4}\int^t_0e^{i[\theta_2(t^\prime)-\theta_0(t^\prime)]}\Big( \frac{[g_1(t^\prime)+g_2(t^\prime)][\partial_{g_2}g_1(t^\prime)+1]}{g_c^2-[g_1(t^\prime)+g_2(t^\prime)]^2}-\frac{[g_1(t^\prime)-g_2(t^\prime)][\partial_{g_2}g_1(t^\prime)-1]}{g_c^2-[g_1(t^\prime)-g_2(t^\prime)]^2} \Big)(-g_c)v(t^\prime) dt^\prime \notag \\
	=&\frac{\sqrt{2}}{4}\int^g_{g_c/k}e^{i[\theta_2(g_2)-\theta_0(g_2)]}\Big( \frac{(g_1+g_2)(\partial_{g_2}g_1+1)}{g_c^2-(g_1+g_2)^2}-\frac{(g_1-g_2)(\partial_{g_2}g_1-1)}{g_c^2-(g_1-g_2)^2} \Big) dg_2 \notag \\
	=&\frac{1}{\sqrt{2}}\int^g_{g_c/k}e^{i\Theta(g_2)}\mathcal{F}(g_2)dg_2,
\end{align}
where the two functions $\mathcal{F}(g_2)$ and $\Theta(g_2)$ are
\begin{align}
	\mathcal{F}(g_2)=&\frac{1}{2}[ \frac{(g_1+g_2)(\partial_{g_2}g_1+1)}{g_c^2-(g_1+g_2)^2}-\frac{(g_1-g_2)(\partial_{g_2}g_1-1)}{g_c^2-(g_1-g_2)^2} ], \\
	\Theta(g_2)=&\theta_2(g_2)-\theta_0(g_2)=-\int^{g_2}_{g_c/k}\frac{2\Delta(g^\prime_2)}{g_cv(g^\prime_2)}dg^\prime_2.
\end{align}
Around the triple point $(g_1,g_2)=(g_c,0)$, function $\mathcal{F}(g_2)$ and the energy gap $\Delta(g_2)$ can be approximated as
\begin{align}
	\mathcal{F}(g_2)=&\frac{(g_c^2+g_1^2-g_2^2)g_2\partial_{g_2}g_1+g_1(g_c^2-g_1^2+g_2^2)}{[(g_c+g_1)^2-g_2^2][(g_c-g_1)^2-g_2^2]} \notag \\
	\approx&\frac{[(1-2k^2)+4k(1-k^2)\frac{g_2}{g_c}](\frac{g_2}{g_c})^2}{g_c[4(4k^2-1)-8k(6k^2-1)\frac{g_2}{g_c}+(68k^4-12k^2+1)(\frac{g_2}{g_c})^2](\frac{g_2}{g_c})^2}
	\approx-\frac{1}{8g_c},  \\
	\Delta(g_2)=&\omega\big[\big(1-[1-(2k+1)\frac{g_2}{g_c}+k^2(\frac{g_2}{g_c})^2]^2\big) \big(1-[1-(2k-1)\frac{g_2}{g_c}+k^2(\frac{g_2}{g_c})^2]^2\big)\big]^{1/2} \notag \\
	\approx&4\omega k\frac{g_2}{g_c}.
\end{align}
Similarly, we can have that
\begin{align}
	\frac{\partial_{g_2}\mathcal{F}(g_2)}{\mathcal{F}(g_2)}
	\approx\frac{5k}{g_c}.
\end{align}
The optimal ramp rate should be set as
\begin{align}
	v(g_2)=\frac{2\delta}{5k}\Delta(g_2)\approx\frac{8\delta\omega}{5}\frac{g_2}{g_c}
\end{align}
with a small parameter $\delta\ll1$, which decreases to zero in the same way with that of energy gap. The excitation probability can be approximated as $\vert c_2(g_2)\vert^2\approx\frac{\delta^2}{25k^2}\ll1$.
The average photon number $N$ and evolution time $T$ can be calculated as
\begin{align}
	N=&\frac{1}{4}[\sqrt{\frac{g^2_c-(g_1+g_2)^2}{g^2_c-(g_1-g_2)^2}}+\sqrt{\frac{g^2_c-(g_1-g_2)^2}{g^2_c-(g_1+g_2)^2}}]-\frac{1}{2}
	\approx\frac{1}{4}(\sqrt{\frac{2k-1}{2k+1}}+\sqrt{\frac{2k+1}{2k-1}})-\frac{1}{2}, \\
	T=&\int^{g_2}_{g_c/k}\frac{1}{-g_cv(g_2^\prime)}dg_2^\prime\approx -\frac{5}{8\delta\omega}\ln (\frac{g_2}{g_c}). 
\end{align}
As QFI around the triple point $(g_1,g_2)=(g_c,0)$ is
\begin{align}
	F_\omega=\frac{1}{8\omega^2}\big( [1-(\frac{g_1+g_2}{g_c})^2]^{-1}-[1-(\frac{g_1-g_2}{g_c})^2]^{-1} \big)^2
	\approx\frac{1}{8\omega^2}\frac{1}{(4k^2-1)^2}(\frac{g_2}{g_c})^{-2},
\end{align}
it can be obtained that
\begin{align}
	F_\omega\approx&\frac{1}{8\omega^2}\frac{1}{(4k^2-1)^2}e^{\frac{16\delta\omega}{5}T},
\end{align}
which is a super-Heisenberg scaling with respect to the adiabatic evolution time $T$.

\subsection{\label{sec:sub}Sub-Heisenberg scaling in adiabatic modulations along a kind of curves}
In this section, we study QFI close to the triple point $(g_1,g_2)=(g_c,0)$ by adapting adiabatically interactions $g_1$ and $g_2$ along a kind of curve in a general adaptive manner
\begin{align}
	\frac{g_1(t)}{g_c}=1-k(\frac{g_2(t)}{g_c})^\beta,~ k>1,
	\frac{g_2(t)}{g_c}=k^{-1/\beta}-\int^t_0 v(t^\prime)dt^\prime.
\end{align}
To ensure that this curve locates in the normal phase, it is demanded that $\beta\in(0,1)$.
It can be easily obtained that
\begin{align*}
	g_c+g_1\pm g_2=g_c[2-k(\frac{g_2}{g_c})^\beta\pm\frac{g_2}{g_c}],~ 
	g_c-g_1\pm g_2=g_c[k\pm(\frac{g_2}{g_c})^{1-\beta}](\frac{g_2}{g_c})^\beta,~ 
	\partial_{g_2}g_1=-k\beta(\frac{g_2}{g_c})^{\beta-1},~\dot{g}_2(t)=-g_cv(t).
\end{align*}
Then, the coefficient of excitations in evolving wave-function becomes
\begin{align}
	c_2(t)=&\frac{1}{\sqrt{2}}\int^t_0e^{i[\theta_2(t^\prime)-\theta_0(t^\prime)]}(-\frac{1}{2})\Big( \frac{[g_1(t^\prime)+g_2(t^\prime)][\dot{g}_1(t^\prime)+\dot{g}_2(t^\prime)]}{g_c^2-[g_1(t^\prime)+g_2(t^\prime)]^2}-\frac{[g_1(t^\prime)-g_2(t^\prime)][\dot{g}_1(t^\prime)-\dot{g}_2(t^\prime)]}{g_c^2-[g_1(t^\prime)-g_2(t^\prime)]^2} \Big) dt^\prime  \notag \\
	=&\frac{\sqrt{2}}{4}\int^t_0e^{i[\theta_2(t^\prime)-\theta_0(t^\prime)]}\Big( \frac{[g_1(t^\prime)+g_2(t^\prime)][\partial_{g_2}g_1(t^\prime)+1]}{g_c^2-[g_1(t^\prime)+g_2(t^\prime)]^2}-\frac{[g_1(t^\prime)-g_2(t^\prime)][\partial_{g_2}g_1(t^\prime)-1]}{g_c^2-[g_1(t^\prime)-g_2(t^\prime)]^2} \Big)(-g_c)v(t^\prime) dt^\prime \notag \\
	=&\frac{\sqrt{2}}{4}\int^g_{g_ck^{-1/\beta}}e^{i[\theta_2(g_2)-\theta_0(g_2)]}\Big( \frac{(g_1+g_2)(\partial_{g_2}g_1+1)}{g_c^2-(g_1+g_2)^2}-\frac{(g_1-g_2)(\partial_{g_2}g_1-1)}{g_c^2-(g_1-g_2)^2} \Big) dg_2 \notag \\
	=&\frac{1}{\sqrt{2}}\int^g_{g_ck^{-1/\beta}}e^{i\Theta^\prime(g_2)}\mathcal{F}^\prime(g_2)dg_2.
\end{align}
Around the triple point $(g_1,g_2)=(g_c,0)$, function $\mathcal{F}^\prime(g_2)$ can be approximated as
\begin{align}
	\mathcal{F}(g_2)=&\frac{1}{2}\Big( \frac{(g_1+g_2)(\partial_{g_2}g_1+1)}{g_c^2-(g_1+g_2)^2}-\frac{(g_1-g_2)(\partial_{g_2}g_1-1)}{g_c^2-(g_1-g_2)^2} \Big) \notag \\
	=&\frac{(g_c^2+g_1^2-g_2^2)g_2\partial_{g_2}g_1+g_1(g_c^2-g_1^2+g_2^2)}{[(g_c+g_1)^2-g_2^2][(g_c-g_1)^2-g_2^2]} \notag \\
	=&\frac{g_c^3[2k(1-\beta)(\frac{g_2}{g_c})^\beta+k^2(2\beta-3)(\frac{g_2}{g_c})^{2\beta}+(\frac{g_2}{g_c})^2+k^3(1-\beta)(\frac{g_2}{g_c})^{3\beta}+k(1-\beta)(\frac{g_2}{g_c})^{2+\beta}]}{g_c^4[4k^2(\frac{g_2}{g_c})^{2\beta}-4(\frac{g_2}{g_c})^2-4k^3(\frac{g_2}{g_c})^{3\beta}+4k(\frac{g_2}{g_c})^{2+\beta}+k^4(\frac{g_2}{g_c})^{4\beta}-2k^2(\frac{g_2}{g_c})^{2+2\beta}+(\frac{g_2}{g_c})^4]} \notag \\
	\approx&\frac{1-\beta}{2k g_c}(\frac{g_2}{g_c})^{-\beta},
\end{align}
and the energy gap $\Delta(g_2)$ is approximately written as
\begin{align}
	\Delta(g_2)=&\omega\big[\big(1-[1-k(\frac{g_2}{g_c})^\beta-\frac{g_2}{g_c}]^2\big) \big(1-[1-k(\frac{g_2}{g_c})^\beta+\frac{g_2}{g_c}]^2\big)\big]^{1/2} \notag \\
	=&\omega\big([(2-k(\frac{g_2}{g_c})^\beta)^2-(\frac{g_2}{g_c})^2][k^2(\frac{g_2}{g_c})^{2\beta}-(\frac{g_2}{g_c})^2]\big)^{1/2} \notag \\
	\approx&2\omega k(\frac{g_2}{g_c})^\beta.
\end{align}
Similarly, we can have that
\begin{align}
	\frac{\partial_{g_2}\mathcal{F}(g_2)}{\mathcal{F}(g_2)}\approx\frac{1}{g_c}[-\beta(\frac{g_2}{g_c})^{-1}+\frac{2-\beta}{k^2}(\frac{g_2}{g_c})^{1-2\beta}+\frac{k\beta(1-2\beta)}{2(1-\beta)}(\frac{g_2}{g_c})^{\beta-1}]
	\approx-\frac{\beta}{g_c}(\frac{g_2}{g_c})^{-1}.
\end{align}
The optimal ramp rate should be set as
\begin{align}
	v(g_2)=\frac{2\delta}{\beta}\frac{g_2}{g_c}\Delta(g_2)\approx\frac{4\delta k\omega}{\beta}(\frac{g_2}{g_c})^{\beta+1}
\end{align}
with a small parameter $\delta\ll1$, which decreases to zero faster than that of energy gap. The excitation probability can be approximated as $\vert c_2(g_2)\vert^2\approx\frac{\delta^2}{8\beta^{2}k^{2/\beta}}\ll1$.
The average photon number $N$ and evolution time $T$ can be calculated as
\begin{align}
	N=&\frac{1}{4}[\sqrt{\frac{g^2_c-(g_1+g_2)^2}{g^2_c-(g_1-g_2)^2}}+\sqrt{\frac{g^2_c-(g_1-g_2)^2}{g^2_c-(g_1+g_2)^2}}]-\frac{1}{2}\approx0, \\
	T=&\int^{g_2}_{g_ck^{-1/\beta}}\frac{1}{-g_cv(g_2^\prime)}dg_2^\prime\approx \frac{1}{4\delta k\omega}(\frac{g_2}{g_c})^{-\beta}. 
\end{align}
As QFI around the triple point $(g_1,g_2)=(g_c,0)$ is
\begin{align}
	F_\omega=\frac{1}{8\omega^2}\big( [1-(\frac{g_1+g_2}{g_c})^2]^{-1}-[1-(\frac{g_1-g_2}{g_c})^2]^{-1} \big)^2
	\approx\frac{1}{8k^4\omega^2}(\frac{g_2}{g_c})^{2(1-2\beta)},
\end{align}
which shows that: when $\beta\in(0,1/2)$, $F_\omega\to0$; when $\beta=1/2$, $F_\omega\approx k^{-4}\omega^{-2}/8$; when $\beta\in(1/2,1)$, $F_\omega$ is divergent.
It can be obtained that for $\beta\in(1/2,1)$
\begin{align}
	F_\omega\approx&\frac{1}{8k^4\omega^2}(4\delta k\omega T)^{2(2-1/\beta)},
\end{align}
which is a sub-Heisenberg scaling with respect to the adiabatic evolution time $T$ because $2-1/\beta<1$.

\section{\label{sec:qfi}Quantum Fisher information close to phase boundaries but far from the triple points}
\begin{figure}[htpb]
	\centering
	\includegraphics[width=0.55\textwidth]{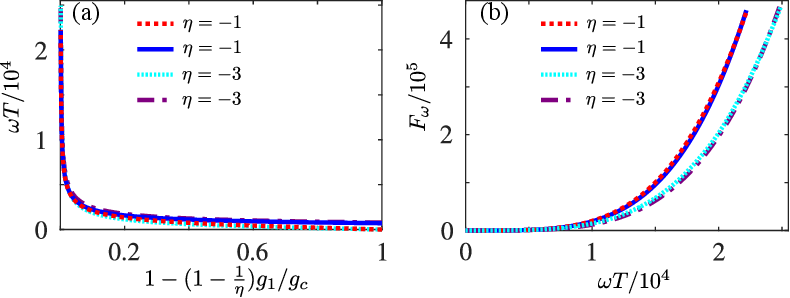}
	\caption{\label{fig:ff2}
	adiabatic evolution time (a) and QFI (b) of the adiabatic evolution paths to final points near phase boundary: $g_1+g_2=g_c$ with $\Omega/\omega=10^6$, $g_c=500$, $\delta=10^{-3}$.
	In figures (a,b), the red-dashed lines and aqua-dashed lines stand for results of real-time adiabatic evolutions with rates $\eta=-1$ and $\eta=-3$ respectively, the blue-solid lines and purple-dot-dashed lines are their corresponding fitted results.
	The fitted function of time is $T=a[1-(1-\frac{1}{\eta})\frac{g_1}{g_c}]^{-1/2}$ with $a=2813$ if $\eta=-1$ and $a=3115$ if $\eta=-3$.
	The fitted function of QFI is $F_\omega=aT^4$ with $a=7.412\times10^{-15}$ for $\eta=-1$ and $a=4.819\times10^{-15}$ for $\eta=-3$.
	In these adiabatic modulations, we vary $g_1$ form $0$ to the final value $0.999(1-\frac{1}{\eta})^{-1}g_c$. 
	}
\end{figure}

To explicitly understand the special roles in this super-Heisenberg scaling played by a triple point, as a comparison, we consider the Heisenberg scaling close to a continuous quantum phase transition.
We employ a simultaneous modulation of both interactions $g_1$ and $g_2$ along a line
\begin{align}
	g_1(t)+\eta g_2(t)=0,~\eta<0,~
	\frac{g_1(t)}{g_c}=\int^t_0\tilde{v}(t^\prime)dt^\prime
\end{align}
to a final point near the phase boundary: $g_1+g_2=g_c$, but far from the triple points, which gives that
\begin{align}
	g_1(t)+g_2(t)=(1-\frac{1}{\eta})g_1(t),~
	g_1(t)-g_2(t)=(1+\frac{1}{\eta})g_1(t),~
	\dot{g}_1(t)=g_c\tilde{v}(t),~\dot{g}_2(t)=-\frac{g_c}{\eta}\tilde{v}(t).
\end{align}
Similarly, the coefficient of excitations in the quasi-adiabatic process is
\begin{align}
	c_2(t)=&\frac{1}{\sqrt{2}}\int^t_0e^{i[\theta_2(t^\prime)-\theta_0(t^\prime)]}(-\frac{1}{2})\Big( \frac{[g_1(t^\prime)+g_2(t^\prime)][\dot{g}_1(t^\prime)+\dot{g}_2(t^\prime)]}{g_c^2-[g_1(t^\prime)+g_2(t^\prime)]^2}-\frac{[g_1(t^\prime)-g_2(t^\prime)][\dot{g}_1(t^\prime)-\dot{g}_2(t^\prime)]}{g_c^2-[g_1(t^\prime)-g_2(t^\prime)]^2} \Big) dt^\prime  \notag \\
	=&-\frac{1}{\sqrt{2}}\int^t_0e^{i[\theta_2(t^\prime)-\theta_0(t^\prime)]}\frac{1}{2g_c}\Big( \frac{(1-\frac{1}{\eta})^2g_1(t^\prime)/g_c}{1-(1-\frac{1}{\eta})^2g^2_1(t^\prime)/g_c^2}-\frac{(1+\frac{1}{\eta})^2g_1(t^\prime)/g_c}{1-(1+\frac{1}{\eta})^2g^2_1(t^\prime)/g^2_c} \Big)g_c\tilde{v}(t^\prime) dt^\prime \notag \\
	=&-\frac{1}{\sqrt{2}}\int^g_0e^{i[\theta_2(g_1)-\theta_0(g_1)]}\frac{1}{2g_c}\Big( \frac{(1-\frac{1}{\eta})^2g_1/g_c}{1-(1-\frac{1}{\eta})^2g^2_1/g_c^2}-\frac{(1+\frac{1}{\eta})^2g_1/g_c}{1-(1+\frac{1}{\eta})^2g^2_1/g^2_c} \Big) dg_1 \notag \\
	=&-\frac{1}{\sqrt{2}}\int^{g}_0e^{i\tilde{\Theta}(g_1)}\tilde{\mathcal{F}}(g_1)dg_1,
\end{align}
in which we have defined two functions $\tilde{\mathcal{F}}(g_1)$ and $\tilde{\Theta}(g_1)$
\begin{align}
	\tilde{\mathcal{F}}(g_1)=&\frac{1}{2g_c}\Big( \frac{(1-\frac{1}{\eta})^2g_1/g_c}{1-(1-\frac{1}{\eta})^2g^2_1/g_c^2}-\frac{(1+\frac{1}{\eta})^2g_1/g_c}{1-(1+\frac{1}{\eta})^2g^2_1/g^2_c} \Big)=-\frac{2g_1\omega^2}{g^2_c\eta\Delta^2(g_1)}, \\
	\tilde{\Theta}(g_1)=&\theta_2(g_1)-\theta_0(g_1)=2\int^{g_1}_0\frac{\Delta(g^\prime)}{g_c\tilde{v}(g^\prime)}dg^\prime
\end{align}
with energy gap $\Delta(g_1)=\omega[1-(1+\frac{1}{\eta})^2(\frac{g_1}{g_c})^2]^{1/2}[1-(1-\frac{1}{\eta})^2(\frac{g_1}{g_c})^2]^{1/2}$.
Around the phase boundary $g_1+g_2=g_c$, the energy gap can be approximated as
\begin{align}
	\Delta(g_1)=\omega[1-(1+\frac{1}{\eta})^2(\frac{g_1}{g_c})^2]^{1/2}[1-(1-\frac{1}{\eta})^2(\frac{g_1}{g_c})^2]^{1/2}
	\approx2\omega\frac{\sqrt{-2\eta}}{1-\eta}[1-(1-\frac{1}{\eta})\frac{g_1}{g_c}]^{1/2},
\end{align}
and in the same way the function $\tilde{\mathcal{F}}(g_1)$ can be approximately written as 
\begin{align}
	\label{eq:mathF}
	\tilde{\mathcal{F}}(g_1)=&-\frac{2g_1\omega^2}{g^2_c\eta\Delta^2(g_1)}
	\approx\frac{1}{4g_c}(1-\frac{1}{\eta})[1-(1-\frac{1}{\eta})\frac{g_1}{g_c}]^{-1},
\end{align}
which can result that
\begin{align}
	\frac{\partial_{g_1}\mathcal{\tilde{F}}(g_1)}{\mathcal{\tilde{F}}(g_1)}=&\frac{1}{g_1}-2\Delta^{-1}(g_1)\partial_{g_1}\Delta(g_1)
	\approx\frac{1}{g_c}(1-\frac{1}{\eta})[1-(1-\frac{1}{\eta})\frac{g_1}{g_c}]^{-1}.
\end{align}
So we can set the optimal ramp rate as
\begin{align}
	\tilde{v}(g_1)=\frac{\delta}{4\omega^2}(1-\eta)\Delta^3(g_1)\approx\frac{4\sqrt{2}\delta\omega}{(1-\eta)^2}(-\eta)^{3/2}[1-(1-\frac{1}{\eta})\frac{g_1}{g_c}]^{3/2}
\end{align}
with a small parameter $\delta\ll1$, which decreases to zero in a way much faster than that of the energy gap. The probability of excitations can be approximated as $\vert c_2(g_1)\vert^2\approx\frac{\delta^2}{32}\ll1$.
The evolution time $T$ and average photon number $N$ can be calculated as
\begin{align}
	\label{eq:time}
	T=&\int^{g_1}_0\frac{1}{g_c\tilde{v}(g)}dg\approx\frac{1-\eta}{2\sqrt{-2\eta}\delta\omega}[1-(1-\frac{1}{\eta})\frac{g_1}{g_c}]^{-1/2}, \\
	N\approx&\frac{1}{4}\sqrt{\frac{1-(1+\frac{1}{\eta})^2(\frac{g_1}{g_c})^2}{1-(1-\frac{1}{\eta})^2(\frac{g_1}{g_c})^2}}
	\approx\frac{\sqrt{-2\eta}}{4(1-\eta)}[1-(1-\frac{1}{\eta})\frac{g_1}{g_c}]^{-1/2}.
\end{align}
As QFI around the phase boundary is
\begin{align}
	F_\omega=&\frac{1}{8\omega^2}\big( [1-(\frac{g_1+g_2}{g_c})^2]^{-1}-[1-(\frac{g_1-g_2}{g_c})^2]^{-1} \big)^2
	\approx\frac{1}{32\omega^2}[1-(1-\frac{1}{\eta})\frac{g_1}{g_c}]^{-2},
\end{align}
it can be obtained that
\begin{align}
	F_\omega\approx2\delta^2N^2T^2\approx2\delta^4\omega^2\frac{\eta^2}{(1-\eta)^4}T^4,
\end{align}
from which we can know that the adiabatic evolution approach can achieve the Heisenberg scaling with respect to both photon number $N$ and time $T$ around the phase boundaries but far from the triple points [see FIG. \ref{fig:ff2}].
From Eq. (\ref{eq:mathF}), we can recognize that due to the existence of a divergent function $\tilde{\mathcal{F}}(g_1)$, it is impossible to devise an adiabatic evolution scheme to restrain excitations as well as the critical slowing down effect when the energy gap closes gradually.

\section{\label{sec:mea}Measurements saturating the quantum Cram\'er-Rao bound}
After studying the scaling laws of measurement precision, we turn to clarify measurements that can saturate the quantum Cram\'er-Rao bound. It has been proposed that photon number ($N$) measuring is a suitable probe in the Rabi-model based quantum metrology \cite{garbe2020phase}.
In a measurement of frequency $\omega$, its precision $\Delta\omega$ is closely related with photon number fluctuation $\Delta N=[\langle n^2\rangle-\langle n\rangle^2]^{1/2}$ with $n=a^\dagger a$.
According to error propagation formula, we can have that
\begin{align}
	\label{eq:dw}
	\Delta\omega=\Delta N/\vert\frac{\partial}{\partial\omega}N\vert.
\end{align}
In the normal phase, photon number and its fluctuation are
\begin{align}
	N=\langle a^\dagger a\rangle=\frac{1}{2}[\cosh(2\gamma)-1],~
	\Delta N=\vert\sinh(2\gamma)\vert/\sqrt{2}.
\end{align}
Substituting these two equations into Eq. (\ref{eq:dw}), it can be gotten that
\begin{align}
	\Delta\omega=\frac{1}{\sqrt{2}\vert\partial_\omega\gamma\vert}=\frac{1}{\sqrt{F_\omega}},
\end{align}
which apparently saturates the Cram\'er-Rao bound. For a measurement, we can define the signal-to-noise ratio (SNR) of frequency $\omega$
\begin{align}
	\label{eq:Some}
	S_{\omega,\psi/\rho}=(\frac{\partial_\omega N}{\Delta N})^2=\frac{(\partial_\omega\langle n\rangle)^2}{\langle n^2\rangle-\langle n\rangle^2}
\end{align}
to identify its corresponding measurement precision, i.e. $\Delta\omega=1/\sqrt{S_{\omega,\psi/\rho}}$. Here, a subscript $\psi$ or $\rho$ is used to indicate the measuring state is pure or mixed.
So only when SNR $S_{\omega,\psi/\rho}$ is identity to the QFI $F_{\omega}$, can we arrive at the highest precision.

\begin{figure}[htbp]
	\centering
	\includegraphics[width=0.65\textwidth]{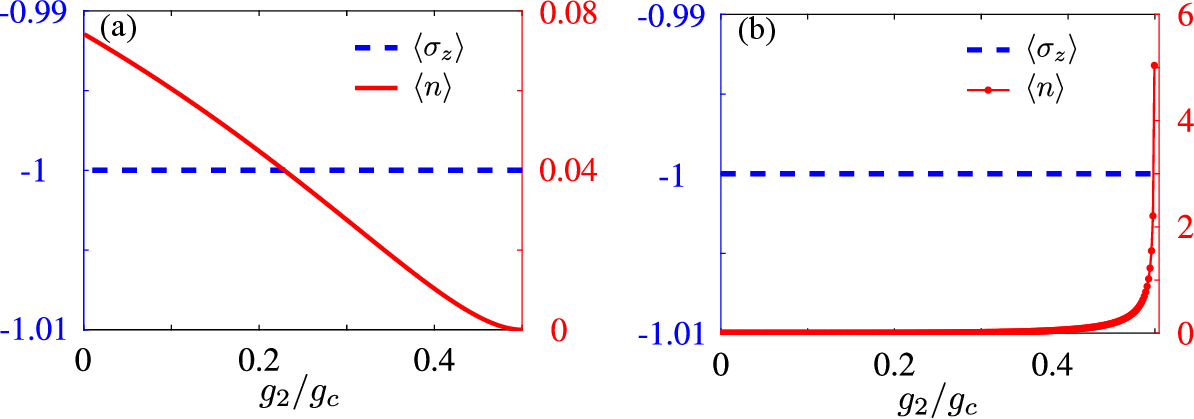}
	\caption{\label{fig:ff5}
	Mean values of photon number $n$ and spin $\sigma_z$ of the instantaneous ground states along adiabatical evolution paths.
	The evolution paths are $g_1+2g_2=g_c$ by varying $g_2$ from $0.5g_c$ to $0.001g_c$ in figure (a) and $g_1-g_2=0$ by varying $g_2$ from $0$ to $0.4995g_c$ in figure (b).
	}
\end{figure}

\section{Effects of dissipation on the adiabatic evolution}
An actual light-atom system will inevitably interact with external environments, which will definitely introduce decoherence and at last destroy the needed adiabatic evolutions in criticality sensing as well as the super-HS.
It is necessary to analyse effects of dissipation in the open aQRM due to photon loss and spin decay. The dissipative dynamics can be described by a master equation
\begin{align}
	\frac{\partial}{\partial t}\rho(t)=-\frac{i}{\hbar}[H,\rho(t)]+\mathcal{K}_{p}\mathcal{L}[a]\rho(t)+\mathcal{K}_{a}\mathcal{L}[\sigma_-]\rho(t)
\end{align}
with dissipative rates $\mathcal{K}_p\ll\omega$, $\mathcal{K}_a\ll\Omega$ and damping superoperator $\mathcal{L}[O]\rho=O\rho O^\dagger-\frac{1}{2}\{O^\dagger O,\rho\}$.
Effects of dissipation on the above adiabatic evolutions can be analysed by directly solving this master equation with Hamiltonian varying along a proposed evolution path to a critical point.
In the process of real-time evolution, effects of dissipation are affected by the populations of atom inner-states and photonic field in the instantaneous ground states of Hamiltonian $H$.
For example, if we approach the triple point $(g_1,g_2)=(g_c,0)$ along a path: $g_1+2g_2=g_c$, as the instantaneous ground states contain very little excitations in spin and photonic field [see FIG. \ref{fig:ff5}(a)], it will remain a long time coherent evolution; 
however, for the path: $g_1-g_2=0$, because there exist many excitations in photonic field near the critical point $(g_1,g_2)=(0.5,0.5)g_c$, dissipation of photon field will lead to decoherence and it becomes difficult to remain coherent evolution.
Hence we can expect good performance of the advised criticality enhanced metrology around a triple point even with existence of dissipation.
We can utilize the SNR $S_{\omega,\rho}$ defined in Eq. (\ref{eq:Some}) to determine it quantificationally.


\section{\label{sec:jcmodel}Super-Heisenberg scaling in the Jaynes-Cummings model with a squeezing bosonic mode}
In this section, we explore another example of triple point criticality that holds super-Heisenberg scaling.
As a derivative model of QRM, the quantum Jaynes-Cummings model (JCM) with a squeezing bosonic mode describes interactions between a two-level system and a single-mode squeezed light field generated through an optical parametric amplification process \cite{qin2018exponentially,tang2022quantum}.
Its effective Hamiltonian can be in the form of
\begin{align}
	\tilde{H}/\hbar=\frac{\tilde{\Omega}}{2}\sigma_z+\tilde{\omega} a^\dagger a+\frac{\tilde{g}}{2}(a^\dagger\sigma_-+a\sigma_+)+\frac{h}{2}(a^{\dagger2}+a^2),
\end{align}
where $\tilde{\Omega}$ is transition frequency of the two-level system with ground state $\vert\downarrow\rangle$ and excited state $\vert\uparrow\rangle$, frequency of the squeezed light field is $\tilde{\omega}$, $\tilde{g}$ represents their rotating-wave coupling strength, $h$ is the strength of two-photon squeezing.
It should be satisfied that $\vert h\vert\le \tilde{\omega}$ for this physical model. For simplicity, we set the interaction to be positive $\tilde{g}>0$.
We first discuss its phase diagram of ground states, then further analyze QFI around its triple point.

\subsection{\label{sec:pdjc}Phase diagram}
We try to investigate ground states in the case of $\tilde{\Omega}\gg\tilde{\omega},\tilde{g},h$. When the interaction $\tilde{g}$ is weak compared with light field frequency $\tilde{\omega}$, its ground state is a normal phase. By making a unitary Schrieffer-Wolff transformation
\begin{align}
	\tilde{U}_n=\exp[-\tilde{\xi}_n],~
	\tilde{\xi}_n=-\tilde{\xi}^\dagger_n=\frac{\tilde{g}(\tilde{\Omega}+\tilde{\omega})}{2(\tilde{\Omega}^2-\tilde{\omega}^2+h^2)}(a\sigma_+-a^\dagger\sigma_-)+\frac{\tilde{g}h}{2(\tilde{\Omega}^2-\tilde{\omega}^2+h^2)}(a\sigma_+-a^\dagger\sigma_-),
\end{align}
we can obtain a transformed Hamiltonian $\tilde{H}_n/\hbar=\tilde{U}^\dagger_n\tilde{H}\tilde{U}_n/\hbar$. To second order of $\tilde{g}$, Hamiltonian $\tilde{H}_n$ reads
\begin{align}
	\tilde{H}_n/\hbar=&\frac{\tilde{\Omega}}{2}\sigma_z+\tilde{\omega} a^\dagger a+\frac{\tilde{g}^2}{8}\big[\frac{\tilde{\Omega}+\tilde{\omega}}{\tilde{\Omega}^2-\tilde{\omega}^2+h^2}[(2a^\dagger a+1)\sigma_z+1]+\frac{h}{\tilde{\Omega}^2-\tilde{\omega}^2+h^2}(a^{\dagger 2}+a^2)\big]+\frac{h}{2}(a^{\dagger 2}+a^2) \notag \\
	=&\frac{\tilde{\Omega}}{2}\sigma_z+\tilde{\omega} a^\dagger a+\frac{\tilde{g}^2}{8\tilde{\Omega}}[(2a^\dagger a+1)\sigma_z+1]+\frac{h}{2}(a^{\dagger 2}+a^2) 
\end{align}
in the infinite frequency ratio limit $\tilde{\Omega}/\tilde{\omega}\to+\infty$.
As the transition frequency $\tilde{\Omega}$ is dominated, this two-level system will lie at its ground state $\vert\downarrow\rangle$, effective Hamiltonian of the light field then becomes 
\begin{align}
	\tilde{H}_{np}/\hbar=\langle\downarrow\vert \tilde{H}_n/\hbar\vert\downarrow\rangle
	\approx&-\frac{\tilde{\Omega}}{2}+\frac{\tilde{g}^2}{8\tilde{\Omega}}+\tilde{\omega} a^\dagger a-\frac{\tilde{g}^2}{8\tilde{\Omega}}(2a^\dagger a+1)+\frac{h}{2}(a^{\dagger 2}+a^2)  \notag \\
	=&-\frac{1}{2}(\tilde{\Omega}+\tilde{\omega})+\frac{\tilde{\omega}}{2}\frac{\tilde{g}^2}{\tilde{g}_c^2}+\frac{\tilde{\omega}}{2}[(1-\frac{\tilde{g}^2}{\tilde{g}^2_c})(2a^\dagger a+1)+\frac{h}{\tilde{\omega}}(a^{\dagger 2}+a^2)]
\end{align}
with $\tilde{g}_c=2\sqrt{\tilde{\Omega}\tilde{\omega}}$. By applying a squeezing transformation
\begin{align}
	\Gamma(\tilde{\gamma})=\exp[\frac{\tilde{\gamma}}{2}(a^2-a^{\dagger 2})]~\text{with}~
	e^{2\tilde{\gamma}}=\sqrt{\frac{1-\frac{\tilde{g}^2}{\tilde{g}_c^2}+\frac{h}{\tilde{\omega}}}{1-\frac{\tilde{g}^2}{\tilde{g}_c^2}-\frac{h}{\tilde{\omega}}}},~
	\tilde{\gamma}=\frac{1}{4}\ln\frac{1-\frac{\tilde{g}^2}{\tilde{g}_c^2}+\frac{h}{\tilde{\omega}}}{1-\frac{\tilde{g}^2}{\tilde{g}_c^2}-\frac{h}{\tilde{\omega}}},
\end{align}
the effective Hamiltonian $\tilde{H}_{np}$ can be diagonalized as
\begin{align}
	\tilde{H}^d_{np}/\hbar=\Gamma^\dagger(\tilde{\gamma}) \tilde{H}_{np}/\hbar\Gamma(\tilde{\gamma})=\tilde{\Delta} a^\dagger a-\frac{1}{2}(\tilde{\Omega}+\tilde{\omega})+\frac{\tilde{\omega}}{2}\frac{\tilde{g}^2}{\tilde{g}_c^2}+\frac{\tilde{\Delta}}{2}
\end{align}
with $\tilde{\Delta}=\tilde{\omega}\sqrt{[1-(\frac{\tilde{g}}{\tilde{g}_c})^2+\frac{h}{\tilde{\omega}}][1-(\frac{\tilde{g}}{\tilde{g}_c})^2-\frac{h}{\tilde{\omega}}]}$.
At a phase boundary, the energy gap closes, which leads to
\begin{align}
	\frac{h}{\tilde{\omega}}=1-(\frac{\tilde{g}}{\tilde{g}_c})^2~\text{or}~
	\frac{h}{\tilde{\omega}}=-1+(\frac{\tilde{g}}{\tilde{g}_c})^2.
\end{align}
In the original frame, ground states of the light field should be
\begin{align}
	\vert\tilde{\psi}_{np}\rangle=\tilde{U}_n\Gamma(\tilde{\gamma})\vert0\rangle=\Gamma(\tilde{\gamma})\vert0\rangle,
\end{align}
because the unitary Schrieffer-Wolff transformation $\tilde{U}_n=1$ in the limit $\tilde{\Omega}/\tilde{\omega}\to+\infty$.

\begin{figure}[htbp]
	\centering
	\includegraphics[width=0.42\textwidth]{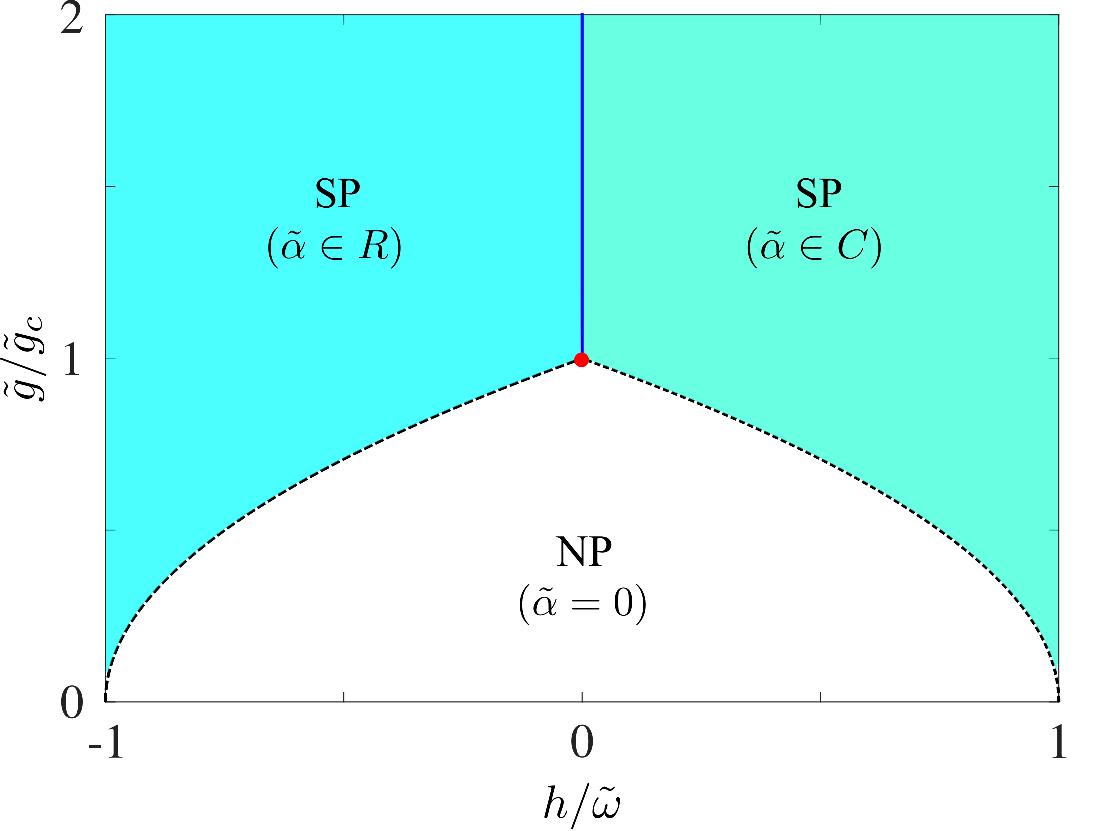}
	\caption{\label{fig:pdjc}
	Phase diagram of the ground states in plane of $\tilde{g}-h$. ``NP" and ``SP" are short for normal phase and superradiant phase respectively. The phase transitions from NP to SP are continuous spontaneously breaking the parity symmetry, while that between SPs turn out to be discontinuous.
	There exist a triple point $(\tilde{g},h)=(g_c,0)$ marked by a red dot.
	}
\end{figure}

When interaction $\tilde{g}$ becomes strong compared with light field frequency $\tilde{\omega}$, the normal phase will turn into superradiant states.
We displace the light field using a displacement transformation $D(\tilde{\alpha})=\exp(\tilde{\alpha} a^\dagger-\tilde{\alpha}^\ast a)$, and the Hamiltonian is transformed into
\begin{align}
	\tilde{H}^\prime/\hbar=D^\dagger(\tilde{\alpha})\tilde{H}D(\tilde{\alpha})/\hbar=&\tilde{\omega}\vert\tilde{\alpha}\vert^2+\frac{h}{2}(\tilde{\alpha}^2+\tilde{\alpha}^{\ast2})+\tilde{\omega}(\tilde{\alpha} a^\dagger+\tilde{\alpha}^\ast a)+h(\tilde{\alpha}^\ast a^\dagger+\tilde{\alpha} a) \notag \\
	&~~~~~+\tilde{\mathcal{H}}_q+\tilde{\omega} a^\dagger a+\frac{\tilde{g}}{2}(a^\dagger\sigma_-+a\sigma_+)+\frac{h}{2}(a^{\dagger2}+a^2)
\end{align}
by using $D^\dagger(\tilde{\alpha})a D(\tilde{\alpha})=a+\tilde{\alpha}$. Concrete form of the new Hamiltonian $\tilde{\mathcal{H}}_q$ is
\begin{align}
	\tilde{\mathcal{H}}_q=\frac{\tilde{\Omega}}{2}\sigma_z+\frac{\tilde{g}}{2}(\tilde{\alpha}^\ast\sigma_-+\tilde{\alpha}\sigma_+)=\frac{\tilde{\Omega}}{2}\sigma_z+\frac{\tilde{g}\vert\tilde{\alpha}\vert}{2}(e^{-i\tilde{\phi}}\sigma_-+e^{i\tilde{\phi}}\sigma_+),
\end{align}
whose eigenvalues are $\tilde{\epsilon}_\pm=\pm\frac{1}{2}\sqrt{\tilde{\Omega}^2+\tilde{g}^2\vert\tilde{\alpha}\vert^2}$. Their corresponding eigenstates can be written as
\begin{align}
	\vert+\rangle=\sin\tilde{\theta}\vert\uparrow\rangle+\cos\tilde{\theta} e^{-i\tilde{\phi}}\vert\downarrow\rangle,~
	\vert-\rangle=\cos\tilde{\theta} e^{i\tilde{\phi}}\vert\uparrow\rangle-\sin\tilde{\theta}\vert\downarrow\rangle,
\end{align}
with $\sin\tilde{\theta}=\frac{1}{\sqrt{2}}\sqrt{1+\frac{\tilde{\Omega}}{\sqrt{\tilde{\Omega}^2+\tilde{g}^2\vert\tilde{\alpha}\vert^2}}}$, $\cos\tilde{\theta}=\frac{1}{\sqrt{2}}\sqrt{1-\frac{\tilde{\Omega}}{\sqrt{\tilde{\Omega}^2+\tilde{g}^2\vert\tilde{\alpha}\vert^2}}}$. It gives that
\begin{align}
	&\vert\uparrow\rangle=\sin\tilde{\theta}\vert+\rangle+\cos\tilde{\theta} e^{-i\tilde{\phi}}\vert-\rangle,~
	\vert\downarrow\rangle=\cos\tilde{\theta} e^{i\tilde{\phi}}\vert+\rangle-\sin\tilde{\theta}\vert-\rangle,\\
	&\sigma_+=\sigma_-^\dagger=\vert\uparrow\rangle\langle\downarrow\vert=\frac{1}{2}\sin(2\tilde{\theta})e^{-i\tilde{\phi}}\tau_z-\sin^2\tilde{\theta}\tau_++\cos^2\tilde{\theta} e^{-i2\tilde{\phi}}\tau_-.
\end{align}
Here we have defined new Pauli operators in the eigen-space
\begin{align}
	\tau_z=\vert+\rangle\langle+\vert-\vert-\rangle\langle-\vert,~
	\tau_+=\tau^\dagger_-=\vert+\rangle\langle-\vert.
\end{align}
In this basis, the transformed Hamiltonian turns out to be
\begin{align}
	\tilde{H}^\prime/\hbar=&\tilde{\omega}\vert\tilde{\alpha}\vert^2+h\vert\tilde{\alpha}\vert^2\cos(2\tilde{\phi})+[(\tilde{\omega}\vert\tilde{\alpha}\vert e^{-i\tilde{\phi}}+h\vert\tilde{\alpha}\vert e^{i\tilde{\phi}}+\frac{\tilde{g}}{4}\sin(2\tilde{\theta})e^{-i\tilde{\phi}}\tau_z)a+\text{h.c.}] \notag \\
	&~~~+\frac{1}{2}\sqrt{\tilde{\Omega}^2+\tilde{g}^2\vert\tilde{\alpha}\vert^2}\tau_z+\tilde{\omega} a^\dagger a+\frac{\tilde{g}}{2}[\cos^2\tilde{\theta}(e^{-i2\tilde{\phi}}a\tau_-+e^{i2\tilde{\phi}}a^\dagger\tau_+)-\sin^2\tilde{\theta}(a^\dagger\tau_-+a\tau_+)]+\frac{h}{2}(a^{\dagger2}+a^2).
\end{align}
It is demanded that
\begin{align}
	\tilde{\omega}\vert\tilde{\alpha}\vert e^{-i\tilde{\phi}}+h\vert\tilde{\alpha}\vert e^{i\tilde{\phi}}-\frac{\tilde{g}}{4}\sin(2\tilde{\theta})e^{-i\tilde{\phi}}=0,
\end{align}
which gives that
\begin{align}
	\begin{cases}
		\tilde{\alpha}=\pm i\frac{\tilde{\Omega}}{\tilde{g}}\sqrt{\frac{\tilde{g}^4}{\tilde{g}^4_c}(1-\frac{h}{\tilde{\omega}})^{-2}-1}, \\
		e^{i2\tilde{\phi}}=e^{-i2\tilde{\phi}}=-1,~\sin^2\tilde{\theta}=\frac{1}{2}[1+\frac{\tilde{g}_c^2}{\tilde{g}^2}(1-\frac{h}{\tilde{\omega}})],
	\end{cases}
\end{align}
or
\begin{align}
	\begin{cases}
		\tilde{\alpha}=\pm \frac{\tilde{\Omega}}{\tilde{g}}\sqrt{\frac{\tilde{g}^4}{\tilde{g}^4_c}(1+\frac{h}{\tilde{\omega}})^{-2}-1}, \\
		e^{i2\tilde{\phi}}=e^{-i2\tilde{\phi}}=1,~\sin^2\tilde{\theta}=\frac{1}{2}[1+\frac{\tilde{g}_c^2}{\tilde{g}^2}(1+\frac{h}{\tilde{\omega}})].
	\end{cases}
\end{align}
When the displacement $\tilde{\alpha}$ is real, Hamiltonian $\tilde{H}^\prime$ becomes
\begin{align}
	\tilde{H}^\prime/\hbar=&\frac{\tilde{\Omega}}{4}[\frac{\tilde{g}^2}{\tilde{g}_c^2}(1+\frac{h}{\tilde{\omega}})^{-1}-\frac{\tilde{g}^2_c}{\tilde{g}^2}(1+\frac{h}{\tilde{\omega}})]+\frac{\tilde{\Omega}^\prime}{2}\tau_z+\tilde{\omega} a^\dagger a+\frac{\tilde{g}_1}{2}(a\tau_++a^\dagger\tau_-)+\frac{\tilde{g}_2}{2}(a^\dagger\tau_++a\tau_-)+\frac{h}{2}(a^{\dagger2}+a^2)
\end{align}
with $\tilde{\Omega}^\prime=\tilde{\Omega}\frac{\tilde{g}^2}{\tilde{g}_c^2}(1+\frac{h}{\tilde{\omega}})^{-1}$, $\tilde{g}_1=-\frac{\tilde{g}}{2}[1+\frac{\tilde{g}^2_c}{\tilde{g}^2}(1+\frac{h}{\tilde{\omega}})]$ and $\tilde{g}_2=\frac{\tilde{g}}{2}[1-\frac{\tilde{g}^2_c}{\tilde{g}^2}(1+\frac{h}{\tilde{\omega}})]$;
when the displacement $\tilde{\alpha}$ is pure complex, the transformed Hamiltonian becomes
\begin{align}
	\tilde{H}^\prime/\hbar=&\frac{\tilde{\Omega}}{4}[\frac{\tilde{g}^2}{\tilde{g}_c^2}(1-\frac{h}{\tilde{\omega}})^{-1}-\frac{\tilde{g}^2_c}{\tilde{g}^2}(1-\frac{h}{\tilde{\omega}})]+\frac{\tilde{\Omega}^\prime}{2}\tau_z+\tilde{\omega} a^\dagger a+\frac{\tilde{g}_1}{2}(a\tau_++a^\dagger\tau_-)+\frac{\tilde{g}_2}{2}(a^\dagger\tau_++a\tau_-)+\frac{h}{2}(a^{\dagger2}+a^2)
\end{align}
with $\tilde{\Omega}^\prime=\tilde{\Omega}\frac{\tilde{g}^2}{\tilde{g}_c^2}(1-\frac{h}{\tilde{\omega}})^{-1}$, $\tilde{g}_1=-\frac{\tilde{g}}{2}[1+\frac{\tilde{g}^2_c}{\tilde{g}^2}(1-\frac{h}{\tilde{\omega}})]$ and $\tilde{g}_2=-\frac{\tilde{g}}{2}[1-\frac{\tilde{g}^2_c}{\tilde{g}^2}(1-\frac{h}{\tilde{\omega}})]$.
We make a unitary Schrieffer-Wolff transformation
\begin{align}
	\tilde{U}_s=\exp[-\tilde{\xi}_s],~
	\tilde{\xi}_s=-\tilde{\xi}^\dagger_s=\frac{\tilde{g}_1(\tilde{\Omega}^\prime+\tilde{\omega})-\tilde{g}_2h}{2(\tilde{\Omega}^{\prime2}-\tilde{\omega}^2+h^2)}(a\tau_+-a^\dagger\tau_-)+\frac{\tilde{g}_1h+\tilde{g}_2(\tilde{\Omega}^\prime-\tilde{\omega})}{2(\tilde{\Omega}^{\prime2}-\tilde{\omega}^2+h^2)}(a^\dagger\tau_+-a\tau_-)
\end{align}
and obtain a transformed Hamiltonian $\tilde{H}_s/\hbar=\tilde{U}^\dagger_s\tilde{H}^\prime \tilde{U}_s/\hbar$. To second order of $\tilde{g}_1$ and $\tilde{g}_2$, Hamiltonian $\tilde{H}_s$ reads
\begin{align}
	\tilde{H}_s/\hbar=&\frac{\tilde{\Omega}^\prime}{2}\tau_z+\tilde{\omega} a^\dagger a+\frac{h}{2}(a^2+a^{\dagger2})+\frac{(\tilde{g}_1^2+\tilde{g}_2^2)\tilde{\Omega}^\prime+(\tilde{g}_1^2-\tilde{g}_2^2)\tilde{\omega}}{8(\tilde{\Omega}^{\prime2}-\tilde{\omega}^2+h^2)}(2a^\dagger a+1)\tau_z \notag \\
	&~~~~~+\frac{2\tilde{g}_1\tilde{g}_2\tilde{\Omega}^\prime+(\tilde{g}_1^2-\tilde{g}_2^2)h}{8(\tilde{\Omega}^{\prime2}-\tilde{\omega}^2+h^2)}(a^{\dagger 2}+a^2)\tau_z+\frac{(\tilde{g}_1^2-\tilde{g}_2^2)\tilde{\Omega}^\prime+(\tilde{g}_1^2+\tilde{g}_2^2)\tilde{\omega}-2\tilde{g}_1\tilde{g}_2h}{8(\tilde{\Omega}^{\prime2}-\tilde{\omega}^2+h^2)} \notag \\
	=&\frac{\tilde{\Omega}^\prime}{2}\tau_z+\tilde{\omega} a^\dagger a+\frac{h}{2}(a^2+a^{\dagger2})+\frac{\tilde{g}_1^2+\tilde{g}_2^2}{8\tilde{\Omega}^{\prime}}(2a^\dagger a+1)\tau_z 
	+\frac{\tilde{g}_1\tilde{g}_2}{4\tilde{\Omega}^{\prime}}(a^{\dagger 2}+a^2)\tau_z+\frac{\tilde{g}_1^2-\tilde{g}_2^2}{8\tilde{\Omega}^{\prime}}
\end{align}
in the limit $\tilde{\Omega}/\tilde{\omega}\to+\infty$, where we have ignored a constant term $\frac{\tilde{\Omega}}{4}\frac{\tilde{g}^2}{\tilde{g}_c^2}(1\pm\frac{h}{\tilde{\omega}})^{-1}[1-\frac{\tilde{g}^4_c}{\tilde{g}^4}(1\pm\frac{h}{\tilde{\omega}})^2]$.
As the two-level system prefers to stay at its ground state $\vert-\rangle$, effective Hamiltonian of the light field becomes
\begin{align}
	\tilde{H}_{sp}/\hbar=\langle-\vert \tilde{H}_s/\hbar\vert-\rangle
	=&-\frac{\tilde{\Omega}^\prime}{2}+\tilde{\omega} a^\dagger a+\frac{h}{2}(a^2+a^{\dagger2})-\frac{\tilde{g}_1^2+\tilde{g}_2^2}{8\tilde{\Omega}^{\prime}}(2a^\dagger a+1)-\frac{\tilde{g}_1\tilde{g}_2}{4\tilde{\Omega}^{\prime}}(a^{\dagger 2}+a^2)+\frac{\tilde{g}_1^2-\tilde{g}_2^2}{8\tilde{\Omega}^{\prime}} \notag \\
	=&-\frac{1}{2}(\tilde{\Omega}^\prime+\tilde{\omega})+\frac{\tilde{\omega}}{2}\frac{\tilde{g}_1^2-\tilde{g}_2^2}{\tilde{g}_c^{\prime2}}+\frac{\tilde{\omega}}{2}[(1-\frac{\tilde{g}_1^2+\tilde{g}_2^2}{\tilde{g}_c^{\prime2}})(2a^\dagger a+1)+(\frac{h}{\tilde{\omega}}-\frac{2\tilde{g}_1\tilde{g}_2}{\tilde{g}_c^{\prime2}})(a^{\dagger 2}+a^2)] 
\end{align}
with $\tilde{g}^\prime_c=2\sqrt{\tilde{\Omega}^\prime\tilde{\omega}}$.
By applying a squeezing transformation
\begin{align}
	\Gamma(\tilde{\gamma}^\prime)=\exp[\frac{\tilde{\gamma}^\prime}{2}(a^2-a^{\dagger 2})]~\text{with}~
	e^{2\tilde{\gamma}^\prime}=\sqrt{\frac{1+\frac{h}{\tilde{\omega}}-\frac{(\tilde{g}_1+\tilde{g}_2)^2}{\tilde{g}^{\prime2}_c}}{1-\frac{h}{\tilde{\omega}}-\frac{(\tilde{g}_1-\tilde{g}_2)^2}{\tilde{g}^{\prime2}_c}}},~
	\tilde{\gamma}^\prime=\frac{1}{4}\ln\frac{1+\frac{h}{\tilde{\omega}}-\frac{(\tilde{g}_1+\tilde{g}_2)^2}{\tilde{g}^{\prime2}_c}}{1-\frac{h}{\tilde{\omega}}-\frac{(\tilde{g}_1-\tilde{g}_2)^2}{\tilde{g}^{\prime2}_c}},
\end{align}
the effective Hamiltonian $\tilde{H}_{sp}$ can be diagonalized as
\begin{align}
	\tilde{H}^d_{sp}/\hbar=\Gamma^\dagger(\tilde{\gamma}^\prime) \tilde{H}_{sp}\Gamma(\tilde{\gamma}^\prime)/\hbar=-\frac{1}{2}(\tilde{\Omega}^\prime+\tilde{\omega})+\frac{\tilde{\omega}}{2}\frac{\tilde{g}_1^2-\tilde{g}_2^2}{\tilde{g}_c^{\prime2}}+\frac{\tilde{\Delta}^\prime}{2}+\tilde{\Delta}^\prime a^\dagger a
\end{align}
with $\tilde{\Delta}^\prime=\tilde{\omega}\sqrt{[1-\frac{h}{\tilde{\omega}}-\frac{(\tilde{g}_1-\tilde{g}_2)^2}{\tilde{g}^{\prime2}_c}][1+\frac{h}{\tilde{\omega}}-\frac{(\tilde{g}_1+\tilde{g}_2)^2}{\tilde{g}^{\prime2}_c}]}$.
So, if the displacement $\tilde{\alpha}$ is real, we can obtain the squeezing factor
\begin{align}
	\tilde{\gamma}^\prime=\frac{1}{4}\ln\frac{1+\frac{h}{\tilde{\omega}}-\frac{\tilde{g}_c^4}{\tilde{g}^{4}}(1+\frac{h}{\tilde{\omega}})^3}{-2\frac{h}{\tilde{\omega}}}
\end{align}
and diagonalized Hamiltonian
\begin{align}
	\tilde{H}^d_{sp}/\hbar=\tilde{\Delta}^\prime a^\dagger a-\frac{\tilde{\omega}}{2}+\frac{\tilde{\Omega}}{4}[\frac{\tilde{g}^2}{\tilde{g}_c^2}(1+\frac{h}{\tilde{\omega}})^{-1}+\frac{\tilde{g}^2_c}{\tilde{g}^2}(1+\frac{h}{\tilde{\omega}})]+\frac{\tilde{\omega}}{2}\frac{\tilde{g}_c^2}{\tilde{g}^2}(1+\frac{h}{\tilde{\omega}})^2+\frac{\tilde{\Delta}^\prime}{2}
\end{align}
with $\tilde{\Delta}^\prime=\tilde{\omega}\sqrt{(-2\frac{h}{\tilde{\omega}})[1+\frac{h}{\tilde{\omega}}-\frac{\tilde{g}_c^4}{\tilde{g}^{4}}(1+\frac{h}{\tilde{\omega}})^3]}$;
if the displacement $\tilde{\alpha}$ is pure complex, it can be known that the squeezing factor should be
\begin{align}
	\tilde{\gamma}^\prime=\frac{1}{4}\ln\frac{2\frac{h}{\tilde{\omega}}}{1-\frac{h}{\tilde{\omega}}-\frac{\tilde{g}_c^4}{\tilde{g}^{4}}(1-\frac{h}{\tilde{\omega}})^3}.
\end{align}
Then the diagonalized Hamiltonian becomes
\begin{align}
	\tilde{H}^d_{sp}/\hbar=\tilde{\Delta}^\prime a^\dagger a-\frac{\tilde{\omega}}{2}+\frac{\tilde{\Omega}}{4}[\frac{\tilde{g}^2}{\tilde{g}_c^2}(1-\frac{h}{\tilde{\omega}})^{-1}+\frac{\tilde{g}^2_c}{\tilde{g}^2}(1-\frac{h}{\tilde{\omega}})]+\frac{\tilde{\omega}}{2}\frac{\tilde{g}_c^2}{\tilde{g}^2}(1-\frac{h}{\tilde{\omega}})^2+\frac{\tilde{\Delta}^\prime}{2}
\end{align}
with $\tilde{\Delta}^\prime=\tilde{\omega}\sqrt{2\frac{h}{\tilde{\omega}}[1-\frac{h}{\tilde{\omega}}-\frac{\tilde{g}_c^4}{\tilde{g}^{4}}(1-\frac{h}{\tilde{\omega}})^3]}$.
In the original frame, ground states of the light field are
\begin{align}
	\vert\tilde{\psi}_{sp}\rangle=D(\tilde{\alpha})\tilde{U}_s\Gamma(\tilde{\gamma}^\prime)\vert0\rangle=D(\tilde{\alpha})\Gamma(\tilde{\gamma}^\prime)\vert0\rangle
\end{align}
in the limit $\tilde{\Omega}/\tilde{\omega}\to+\infty$.
Therefore, when $1+\frac{h}{\tilde{\omega}}<\frac{\tilde{g}^2}{\tilde{g}_c^2}$ and $h<0$, $\tilde{\alpha}$ is real; when $1-\frac{h}{\tilde{\omega}}<\frac{\tilde{g}^2}{\tilde{g}_c^2}$ and $h>0$, $\tilde{\alpha}$ is pure complex.
There exist two kinds of superradiant phases in its phase diagram and the phase transition between them is of first-order, in addition, phase transitions from normal phase to superradiant phases are second-order. A triple point $(\tilde{g},h)=(\tilde{g}_c,0)$ appears in the phase diagram [see FIG. \ref{fig:pdjc}].

\subsection{Quantum metrology around the triple point}
We are interested in quantum metrology around the triple point $(\tilde{g},h)=(\tilde{g}_c,0)$. To estimating the light field frequency $\tilde{\omega}$, we adiabatically varying the interaction $\tilde{g}$ and  squeezing strength $h$ along a path in normal phase to a final point near the triple point.
Precision of measuring light-field frequency $\tilde{\omega}$ is bounded by quantum Cram\'er-Rao bound: $\Delta\tilde{\omega}\ge1/\sqrt{\nu \tilde{F}_{\tilde{\omega}}}$, where $\nu$ is the number of independent measurements and $\tilde{F}_{\tilde{\omega}}$ is the QFI relative to parameter $\tilde{\omega}$.
Because the final state is a normal state, the QFI can be computed exactly as $\tilde{F}_{\tilde{\omega}}=4[ \langle\partial_{\tilde{\omega}}\tilde{\psi}_{np}\vert\partial_{\tilde{\omega}}\tilde{\psi}_{np}\rangle-\vert\langle\partial_{\tilde{\omega}}\tilde{\psi}_{np}\vert\tilde{\psi}_{np}\rangle\vert^2 ]=2(\frac{\partial\tilde{\gamma}}{\partial\tilde{\omega}})^2$. Its concrete form is
\begin{align}
	\tilde{F}_{\tilde{\omega}}=\frac{1}{2\tilde{\omega}^2}[\frac{\frac{h}{\tilde{\omega}}}{(1-\frac{\tilde{g}^2}{\tilde{g}_c^2}+\frac{h}{\tilde{\omega}})(1-\frac{\tilde{g}^2}{\tilde{g}_c^2}-\frac{h}{\tilde{\omega}})}]^2,
\end{align}
from which it can be learned that at the critical point $(\tilde{g},h)=(\tilde{g}_c,0)$, the QFI $\tilde{F}_{\tilde{\omega}}$ can be finite or divergent depending on the adiabatic evolution path.
For example, near the triple point $(\tilde{g},h)=(\tilde{g}_c,0)$, we assume that $1-\frac{\tilde{g}^2}{\tilde{g}^2_c}=\tilde{k}(\frac{h}{\tilde{\omega}})^{\tilde{\beta}} (\tilde{k}>1, 0<\tilde{\beta}\le1)$.
If $0<\tilde{\beta}\le\frac{1}{2}$, we obtain a finite QFI $\tilde{F}_{\tilde{\omega}}\approx \frac{1}{2\tilde{\omega}^2\tilde{k}^4}(\frac{h}{\tilde{\omega}})^{2(1-2\tilde{\beta})}$; when $\frac{1}{2}<\tilde{\beta}<1$, the QFI $\tilde{F}_{\tilde{\omega}}\approx \frac{1}{2\tilde{\omega}^2\tilde{k}^4}(\frac{h}{\tilde{\omega}})^{2(1-2\tilde{\beta})}\to\infty$; if $\tilde{\beta}=1$, the QFI $\tilde{F}_{\tilde{\omega}}\approx \frac{1}{2\tilde{\omega}^2(\tilde{k}^2-1)^2}(\frac{h}{\tilde{\omega}})^{-2}\to\infty$.
We next inspect relationships between the QFI $\tilde{F}_{\tilde{\omega}}$, evolution time $T$ and average photon number $N$ used in the critical sensing.

\begin{figure}[htpb]
	\centering
	\includegraphics[width=0.47\textwidth]{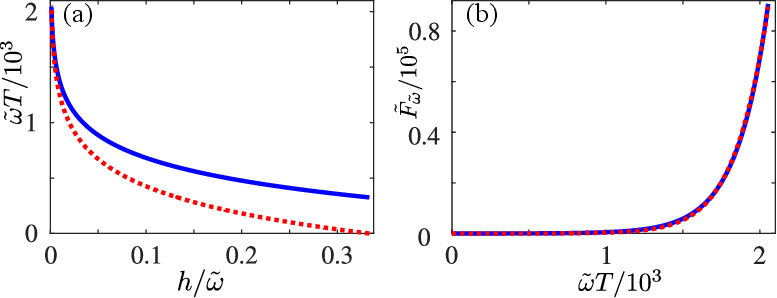}
	\caption{\label{fig:ff7}
	adiabatic evolution time (a) and QFI (b) of the adiabatic evolution path to a final point near the triple point: $(\tilde{g},h)=(\tilde{g}_c,0)$ with $\tilde{\Omega}/\tilde{\omega}=10^6$, $\tilde{g}_c=500$, $\delta=10^{-3}$.
	In figures (a,b), the red-dashed lines stand for results of real-time adiabatic evolutions with parameters $\tilde{k}=3$ and $\beta=1$, the blue-solid lines are their corresponding fitted results.
	The fitted function of time is $T=1084\ln(\frac{h}{\tilde{\omega}})^{-1}$.
	The fitted function of QFI is $\tilde{F}_{\tilde{\omega}}=0.035e^{1.2\times10^{-3}T}$.
	In this adiabatic modulations, we vary $h$ form $\tilde{\omega}/\tilde{k}$ to the final value $10^{-3}\tilde{\omega}$. 
	}
\end{figure}

We adapt adiabatically interactions $\tilde{g}$ and $h$ along the curve using a general adaptive manner
\begin{align}
	1-\frac{\tilde{g}^2(t)}{\tilde{g}^2_c}=\tilde{k}[\frac{h(t)}{\tilde{\omega}}]^{\tilde{\beta}},~ \tilde{k}>1, \frac{1}{2}<\tilde{\beta}\le1,~
	\frac{h(t)}{\tilde{\omega}}=\tilde{k}^{-1/\tilde{\beta}}-\int^t_0 \tilde{v}(t^\prime)dt^\prime,
\end{align}
from which it can be obtained that
\begin{align*}
	1-\frac{\tilde{g}^2}{\tilde{g}^2_c}=\tilde{k}(\frac{h}{\tilde{\omega}})^{\tilde{\beta}},~
	\frac{\dot{h}(t)}{\tilde{\omega}}=-\tilde{v}(t),~
	\frac{\dot{\tilde{g}}(t)}{\tilde{g}_c}=\tilde{k}\tilde{\beta}\frac{\tilde{g}_c}{2\tilde{g}}[\frac{h(t)}{\tilde{\omega}}]^{\tilde{\beta}-1}\tilde{v}(t).
\end{align*}
In this adiabatic process, the instantaneous eigenstates and eigenvalues are
\begin{align}
	\tilde{H}_{np}(t)\vert\tilde{\psi}_n(t)\rangle=\hbar\tilde{\omega}_n(t)\vert\tilde{\psi}_n(t)\rangle,~
	\vert\tilde{\psi}_n(t)\rangle=\Gamma[\tilde{\gamma}(t)]\vert n\rangle
\end{align}
with $n=0,1,2,\cdots$, instantaneous frequency $\tilde{\omega}_n(t)=n\tilde{\Delta}(t)=n\tilde{\omega}\sqrt{[1-(\frac{\tilde{g}(t)}{\tilde{g}_c})^2+\frac{h(t)}{\tilde{\omega}}][1-(\frac{\tilde{g}(t)}{\tilde{g}_c})^2-\frac{h(t)}{\tilde{\omega}}]}$. We have shifted the energy zero point by ignoring terms not including $\tilde{\Delta}(t)$.
Starting from the initial state $\vert\tilde{\psi}(0)\rangle=\vert\tilde{\psi}_{n}(0)\rangle=\vert\tilde{\psi}_{np}\rangle$, at time $t$ its state $\vert\tilde{\psi}(t)\rangle$ can be expanded as
\begin{align}
	\vert\tilde{\psi}(t)\rangle=\sum_{n}\tilde{c}_n(t)e^{-i\tilde{\theta}_n(t)}\vert\tilde{\psi}_n(t)\rangle
\end{align}
with dynamical phase $\tilde{\theta}_n(t)=\int^t_0\tilde{\omega}_n(t^\prime)dt^\prime$.
Using time-dependent perturbation theory, the coefficient  $\tilde{c}_n(t)$ is given by
\begin{align}
	\tilde{c}_n(t)=-\int^t_0e^{i[\tilde{\theta}_n(t^\prime)-\tilde{\theta}_0(t^\prime)]}\langle\psi_m(t^\prime)\vert\frac{\partial}{\partial t^\prime}\psi_0(t^\prime)\rangle dt^\prime
	=\frac{1}{\sqrt{2}}\int^t_0e^{i[\tilde{\theta}_n(t^\prime)-\tilde{\theta}_0(t^\prime)]}[\frac{\partial}{\partial t^\prime}\tilde{\gamma}(t^\prime)]\delta_{n,2} dt^\prime,
\end{align}
so only the transition to excitations with $n=2$ is important and we can have that
\begin{align}
	\tilde{c}_2(t)=&\frac{1}{\sqrt{2}}\int^t_0e^{i[\tilde{\theta}_2(t^\prime)-\tilde{\theta}_0(t^\prime)]}[\frac{\partial}{\partial t^\prime}\tilde{\gamma}(t^\prime)] dt^\prime \notag \\
	=&\frac{1}{\sqrt{2}}\int^t_0e^{i[\tilde{\theta}_2(t^\prime)-\tilde{\theta}_0(t^\prime)]}\frac{1}{4}[\frac{\tilde{k}\tilde{\beta}(\frac{h}{\tilde{\omega}})^{\tilde{\beta}-1}-1}{1-\frac{\tilde{g}^2}{\tilde{g}_c^2}-\frac{h}{\tilde{\omega}}}-\frac{\tilde{k}\tilde{\beta}(\frac{h}{\tilde{\omega}})^{\tilde{\beta}-1}+1}{1-\frac{\tilde{g}^2}{\tilde{g}_c^2}+\frac{h}{\tilde{\omega}}}]\tilde{v}(t^\prime) dt^\prime \notag \\
	=&\frac{1}{\sqrt{2}}\int^{\frac{h}{\tilde{\omega}}}_{\tilde{k}^{-1/\tilde{\beta}}}e^{i[\tilde{\theta}_2(\frac{h^\prime}{\tilde{\omega}})-\tilde{\theta}_0(\frac{h^\prime}{\tilde{\omega}})]}\frac{1}{4}[\frac{\tilde{k}\tilde{\beta}(\frac{h^\prime}{\tilde{\omega}})^{\tilde{\beta}-1}+1}{\tilde{k}(\frac{h^\prime}{\tilde{\omega}})^{\tilde{\beta}}+\frac{h^\prime}{\tilde{\omega}}}-\frac{\tilde{k}\tilde{\beta}(\frac{h^\prime}{\tilde{\omega}})^{\tilde{\beta}-1}-1}{\tilde{k}(\frac{h^\prime}{\tilde{\omega}})^{\tilde{\beta}}-\frac{h^\prime}{\tilde{\omega}}}]d\frac{h^\prime}{\tilde{\omega}} \notag \\
	=&\frac{1}{\sqrt{2}}\int^{\frac{h}{\tilde{\omega}}}_{\tilde{k}^{-1/\tilde{\beta}}}e^{i\tilde{\Theta}^\prime(\frac{h^\prime}{\tilde{\omega}})}\tilde{\mathcal{F}}^\prime(\frac{h^\prime}{\tilde{\omega}})d\frac{h^\prime}{\tilde{\omega}},
\end{align}
where functions $\tilde{\mathcal{F}}^\prime(\frac{h}{\tilde{\omega}})$ and $\tilde{\Theta}^\prime(\frac{h}{\tilde{\omega}})$ are defined as
\begin{align}
	\tilde{\mathcal{F}}^\prime(\frac{h}{\tilde{\omega}})=&\frac{1}{4}[\frac{\tilde{k}\tilde{\beta}(\frac{h}{\tilde{\omega}})^{\tilde{\beta}-1}+1}{\tilde{k}(\frac{h}{\tilde{\omega}})^{\tilde{\beta}}+\frac{h}{\tilde{\omega}}}-\frac{\tilde{k}\tilde{\beta}(\frac{h}{\tilde{\omega}})^{\tilde{\beta}-1}-1}{\tilde{k}(\frac{h}{\tilde{\omega}})^{\tilde{\beta}}-\frac{h}{\tilde{\omega}}}]
	=\frac{1}{2}\frac{\tilde{k}(1-\tilde{\beta})(\frac{h}{\tilde{\omega}})^{\tilde{\beta}}}{\tilde{k}^2(\frac{h}{\tilde{\omega}})^{2\tilde{\beta}}-\frac{h^2}{\tilde{\omega}^2}}, \\
	\tilde{\Theta}^\prime(\frac{h}{\tilde{\omega}})=&\tilde{\theta}_2(\frac{h}{\tilde{\omega}})-\tilde{\theta}_0(\frac{h}{\tilde{\omega}})=\int^{\frac{h}{\tilde{\omega}}}_{\tilde{k}^{-1/\tilde{\beta}}}\frac{2\tilde{\Delta}(\frac{h^\prime}{\tilde{\omega}})}{-\tilde{v}(\frac{h^\prime}{\tilde{\omega}})}d\frac{h^\prime}{\tilde{\omega}}
\end{align}
with energy gap $\tilde{\Delta}(\frac{h}{\tilde{\omega}})=\tilde{\omega}\sqrt{[\tilde{k}(\frac{h}{\tilde{\omega}})^{\tilde{\beta}}-\frac{h}{\tilde{\omega}}][\tilde{k}(\frac{h}{\tilde{\omega}})^{\tilde{\beta}}+\frac{h}{\tilde{\omega}}]}$.
So around the triple point $(\tilde{g},h)=(\tilde{g}_c,0)$, $\tilde{\mathcal{F}}^\prime(\frac{h}{\tilde{\omega}})=0$ for $\tilde{\beta}=1$, which leads to $\tilde{c}_2(t)=0$. At first sight, it seems that no excitations are stimulated, and the adiabatic evolution can be easily guaranteed even if the slow ramp rates $\tilde{v}(\frac{h}{\tilde{\omega}})$ are set finite and not smaller than the energy gap.
However, time-dependent perturbations require that the slow ramp rates should not exceed related energy gaps.
The ramp rate may be set as
\begin{align}
	\tilde{v}(\frac{h}{\tilde{\omega}})=\delta\tilde{\Delta}(\frac{h}{\tilde{\omega}})=\delta\tilde{\omega}(\tilde{k}^2-1)^{1/2}(\frac{h}{\tilde{\omega}})
\end{align}
with a small parameter $\delta\ll1$.
Around the triple point $(\tilde{g},h)=(\tilde{g}_c,0)$, the average photon number $N$ and evolution time $T$ can be calculated as
\begin{align}
	N=&\frac{1}{4}[\sqrt{\frac{\tilde{k}+1}{\tilde{k}-1}}+\sqrt{\frac{\tilde{k}-1}{\tilde{k}+1}}]-\frac{1}{2}, \\
	T=&\int^{\frac{h}{\tilde{\omega}}}_{\tilde{k}^{-1/\tilde{\beta}}}\frac{1}{-\tilde{v}(\frac{h^\prime}{\tilde{\omega}})}d\frac{h^\prime}{\tilde{\omega}}\approx -\frac{1}{\delta\tilde{\omega}(\tilde{k}^2-1)^{1/2}}\ln(\frac{h}{\tilde{\omega}}), 
\end{align}
from which we can acquire a super-HS as follows [see FIG. \ref{fig:ff7}]
\begin{align}
	\tilde{F}_{\tilde{\omega}}\approx\frac{1}{2\tilde{\omega}^2(\tilde{k}^2-1)^2}e^{2\delta(\tilde{k}^2-1)^{1/2}\tilde{\omega}T}.
\end{align}

For $\frac{1}{2}<\tilde{\beta}<1$, $\tilde{\mathcal{F}}^\prime(\frac{h}{\tilde{\omega}})\approx \frac{1-\tilde{\beta}}{2\tilde{k}}(\frac{h}{\tilde{\omega}})^{-\tilde{\beta}}\to\infty$.
The optimal ramp rate should be set as
\begin{align}
	\tilde{v}(\frac{h}{\tilde{\omega}})=\frac{2\delta}{\tilde{\beta}}\frac{h}{\tilde{\omega}}\tilde{\Delta}(\frac{h}{\tilde{\omega}})\approx\frac{2\delta\tilde{k}\tilde{\omega}}{\tilde{\beta}}(\frac{h}{\tilde{\omega}})^{\tilde{\beta}+1}
\end{align}
with a small parameter $\delta\ll1$, and the excitation probability can be approximated as $\vert \tilde{c}_2\vert^2\approx\frac{\delta^2}{8\tilde{\beta}^2}\tilde{k}^{-2/\tilde{\beta}}\ll1$.
Around the triple point $(\tilde{g},h)=(\tilde{g}_c,0)$, the average photon number $N$ and evolution time $T$ can be calculated as
\begin{align}
	N=&\frac{1}{4}[\sqrt{\frac{1-\frac{\tilde{g}^2}{\tilde{g}_c^2}+\frac{h}{\tilde{\omega}}}{1-\frac{\tilde{g}^2}{\tilde{g}_c^2}-\frac{h}{\tilde{\omega}}}}+\sqrt{\frac{1-\frac{\tilde{g}^2}{\tilde{g}_c^2}-\frac{h}{\tilde{\omega}}}{1-\frac{\tilde{g}^2}{\tilde{g}_c^2}+\frac{h}{\tilde{\omega}}}}]-\frac{1}{2}, \\
	T=&\int^{\frac{h}{\tilde{\omega}}}_{\tilde{k}^{-1/\tilde{\beta}}}\frac{1}{-\tilde{v}(\frac{h^\prime}{\tilde{\omega}})}d\frac{h^\prime}{\tilde{\omega}}\approx \frac{1}{2\delta\tilde{\omega}\tilde{k}}(\frac{h}{\tilde{\omega}})^{-\tilde{\beta}}, 
\end{align}
thus it can be obtained that
\begin{align}
	\tilde{F}_{\tilde{\omega}}\approx\frac{1}{2\tilde{\omega}^2\tilde{k}^4}(2\delta\tilde{k}\tilde{\omega}T)^{2(2-1/\tilde{\beta})}.
\end{align}
It is a sub-Heisenberg scaling as $2-1/\tilde{\beta}<1$.

\bibliography{qmSI}